\documentclass[useAMS]{mn2e}
\usepackage{amssymb,amsmath}
\usepackage{graphicx}

\title[Double-Peaked Emitter 3C390.3]
{Correlation between Line Width and Line Flux of     
     Double-Peaked Broad H$\alpha$ of 3C390.3}
\author[Zhang X.-G.]
       {Xue-Guang Zhang$^{1,2}$\thanks{xgzhang@pmo.ac.cn}\\
       $^1$Purple Mountain Observatory, Chinese Academy of Sciences,
             2 Beijing XiLu, NanJing, JiangSu, 210008, P. R. China \\
       $^2$Chinese Center for Antarctic Astronomy, NanJing, 
             JiangSu, 210008, P. R. China}

\date{}
\def\LaTeX{L\kern-.36em\raise.3ex\hbox{a}\kern-.15em
    T\kern-.1667em\lower.7ex\hbox{E}\kern-.125emX}

\begin{document}
\pagerange{\pageref{firstpage}--\pageref{lastpage}} \pubyear{2012}
\maketitle
\label{firstpage}

\begin{abstract}
    In this manuscript, we carefully check the correlation between 
the line width (second moment) and the line flux of the double-peaked 
broad H$\alpha$ of the well-known mapped AGN 3C390.3, in order to 
show some further distinctions between double-peaked emitters and 
normal broad line AGN. Based on the Virialization assumption 
$M_{BH}\propto R_{BLR}\times V^2(BLR)$ and the empirical relation 
$R_{BLR}\propto L^{\sim0.5}$, one strong negative correlation between 
the line width and the line flux of the double-peaked broad lines 
should be expected for 3C390.3, such as the negative correlation 
confirmed for the mapped broad line object NGC5548,  
$R_{BLR}\times V^2(BLR)\propto L^{\sim0.5}\times \sigma^2 = constant$. 
But, based on the public spectra around 1995 from the AGNWATCH 
project for 3C390.3, one reliable positive correlation is found  
between the line width and the line flux of the double-peaked broad 
H$\alpha$. In the context of the proposed theoretical accretion 
disk model for double-peaked emitters, the unexpected positive 
correlation can be naturally explained, due to different time 
delays for inner parts and outer parts of disk-like BLR of 3C390.3. 
Moreover, the Virialization assumption is checked and found to be 
still available for 3C390.3. However, time-varying size of the BLR 
of 3C390.3 can not be expected by the empirical relation 
$R_{BLR}\propto L^{\sim0.5}$. In other words, the mean size of 
BLR of 3C390.3 can be estimated by the continuum luminosity 
(line luminosity), while, the continuum emission strengthening 
leads to the size of BLR decreasing (not increasing) in different 
moments for 3C390.3. Then, we compared our results of 3C390.3 with 
the previous results reported in the literature for the 
other double-peaked emitters, and found that before to clearly correct 
effects from disk physical parameters varying (such as effects of disk 
precession) for long-term observed line spectra, it is not so meaningful 
to discuss the correlation of line parameters of double-peaked broad 
lines. Furthermore, due to the probable 'external' ionizing source 
with so far unclear structures, it is hard to give one conclusion 
that the positive correlation between the line width and the line 
flux can be found for all double-peaked emitters, even after the 
considerations of disk physical parameters varying. However, once 
one positive correlation of broad line parameters was found, the 
accretion disk origination of the broad line should be firstly 
considered.
\end{abstract}


\section{Introduction}

   As one well-known double-peaked emitter (AGN with double-peaked 
broad low-ionization emission lines), 3C390.3 has been studied for 
more than two decades, since clear double-peaked broad balmer lines 
were reported by  Burbidge \& Burbidge (1971, spectrogram of 
3C390.3) and by Perez et al. (1988, clear spectral lines). Based 
on double-peaked appearance of the broad lines and characters of 
long-term variabilities of the broad lines and continuum emission, 
among those proposed theoretical models for double-peaked 
emitters (such as binary black hole model: Begelman et al. 1980, 
Boroson \& Lauer 2009, Gaskell 1996, Lauer \& Boroson 2009, 
Zhang et al. 2007, double stream model: Vellieux \& Zheng 1991, 
Zheng et al. 1990, 1991,  accretion disk model: Bachev 1999, 
Chen et al. 1989, Chen \& Halpern 1989, Chornock et al. 2010, 
Eracleous et al. 1995,  Flohic \& Eracleous 2008, Gaskell 2010, 
Gezari et al. 2007, Hartnoll \& Blackman 2000, 2002, Karas et al. 2001,  
Lewis et al. 2010, Oke 1987, Perez et al. 1988, Storchi-Bergmann, et al., 
2003, Schimoia et al. 2012, Tran 2010), the accretion disk model 
(double-peaked broad lines coming from central accretion disk) has 
been widely accepted for the double-peaked emitter 3C390.3 
(Chen et al. 1989, Chen \& Halpern 1989,  Dietrich et al. 2012, 
Eracleous et al. 1997, Livio \& Xu 1997, Oke 1987, Perez et al. 1988, 
Popovic et al. 2011, Shapovalova et al. 2001, 2010, Zhang 2011a). 
Then based on the proposed theoretical accretion disk model, besides 
the different line profiles, we try to find some interesting 
distinctions between the double-peaked emitter 3C390.3 and 
normal broad line AGN, which is the main objective of our 
manuscript.

     As we commonly know that motions of broad emission line 
clouds in BLR (broad emission line regions) of AGN are gravitationally 
dominated by central mass of AGN (Gaskell 1988, 1996, Krause et al. 2011, 
Kollatschny \& Zetzl1 2011, Netzer \& Marziani 2010, 
Peterson \& Wandel 1999, Wandel et al. 1999, and references therein), 
which provides strong evidence for applying Virialization assumption 
to estimate virial black hole masses of broad line AGN (
Bennert et al. 2011, Collin et al. 2006, Greene \& Ho 2004, 2005, 
Kelly \& Bechtold 2007, Marziani et al. 2003, Netzer \& Marziani 2010, 
Onken et al. 2004, Park et al. 2012, Peterson et al. 2004, Peterson 2010, 
Rafiee \& Hall 2011, Shen \& Liu 2012, Sulentic et al., 2000, 
Vestergaard 2002, Wu et al. 2004). Then based on the Virialization 
assumption, combining with widely accepted empirical relation to 
estimate size of BLR of AGN ($R_{BLR}\propto L^{\sim0.5}$, 
Bentz et al. 2006, 2009, Denney et al. 2010, Kaspi et al. 1996, 2000, 
2005, Greene et al. 2010, Wang \& Zhang 2003), we will have 
\begin{equation}
\begin{split}
M_{BH}&\propto R_{BLR}\times V(BLR)^2\\
&\propto L(H\alpha_B)^{\sim0.5}\times V^2(H\alpha_B)
\end{split}
\end{equation}
where index $H\alpha_B$ means corresponding line parameters of broad 
H$\alpha$. Moreover, we can find that for individual object 
($M_{BH}=constant$ and $L(H\alpha_B)\propto flux(H\alpha_B)$),
\begin{equation}
\begin{split}
&L(H\alpha_B) \times V(H\alpha_B)^4 = constant \\
&flux(H\alpha_B) \times V(H\alpha_B)^4 = constant \\
&L(H\alpha_B) (or\ R_{BLR}\ or\ flux(H\alpha_B))\Uparrow\ \Longrightarrow\ V(H\alpha_B)\Downarrow
\end{split}
\end{equation}
In other words, line width of broad emission lines is decreasing with 
line luminosity (size of BLR, continuum luminosity) increasing. The 
strong expected negative correlation, $R_{BLR}$ versus $V$, have been 
found and reported for several well mapped objects, such as NGC5548 
(Bentz et al. 2007, Denney et al. 2010, Peterson et al. 2002, 2004, 
Wanders \& Peterson 1996), NGC3783 and NGC7469 (Peterson et al. 2004), 
Arp151 (Park et al. 2012). Certainly, the correlation between $R_{BLR}$ 
and $V$ for 3C390.3 is also shown in Peterson et al. (2004), however, 
there is no one reliable correlation confirmed for 3C390.3, due to 
the less number of data points (only 4 data points for 3C390.3 
shown in Figure 8\ in Peterson et al. 2004). In the paper, we will 
carefully re-check the correlation for 3c390.3, because among the 
well-mapped objects, 3c390.3 is the unique double-peaked emitter. 

   In the paper, we check correlation between line width and 
line flux, rather than to check correlation between size of BLR 
and line width, in order to obtain abundant data points. This paper 
is organized as follows. Section 2 gives our data procedures and 
main results. Section 3 shows our discussions and 
conclusions. Here, cosmological parameters 
$H_{0}=70{\rm km\cdot s}^{-1}{\rm Mpc}^{-1}$, $\Omega_{\Lambda}=0.7$ 
and $\Omega_{m}=0.3$ have been adopted.

\section{Main Results}

   All public spectra of 3C390.3 can be collected from the AGNWATCH 
project (http://www.astronomy.ohio-state.edu/\~{}agnwatch/), which is 
a consortium of astronomers who have studied the inner structures of 
AGN through continuum and emission-line variability. Detailed 
descriptions about observational instruments, techniques and 
procedures to reduce the spectra of 3C390.3 can be found in 
Dietrich et al. (1998), which will not be described any more. 
Here, we mainly show our procedures to measure line parameters 
(line width and line flux) of the double-peaked broad lines of 
3C390.3, and show our results about correlation of the line parameters.

\subsection{Line Parameters}

    We mainly focus on line parameters of broad H$\alpha$ of 
3C390.3, due to the following two main reasons. On the one 
hand, the broad H$\alpha$ is much stronger than the broad 
H$\beta$, which leads to more reliable line parameters. On the 
other hand, there is much less contamination on the broad 
H$\alpha$. Only the narrow emission lines, the narrow H$\alpha$ 
and [N~{\sc ii}] doublet, have effects on the measured line 
parameters of the double-peaked broad H$\alpha$. However, for 
the double-peaked broad H$\beta$, besides the common narrow 
H$\beta$ and [O~{\sc iii}] doublet, the extended [O~{\sc iii}] 
doublet and some weak optical Fe~{\sc ii} emission lines 
(Dietrich et al. 2012) have effects on the measured line 
parameters of the broad H$\beta$. Thus, focusing on the 
observed spectrum around H$\alpha$ of 3C390.3 should lead 
to more accurate line parameters and to find more reliable 
correlation of the line parameters.

\begin{figure*}
\centering\includegraphics[height = 22.5cm,width = 18cm]{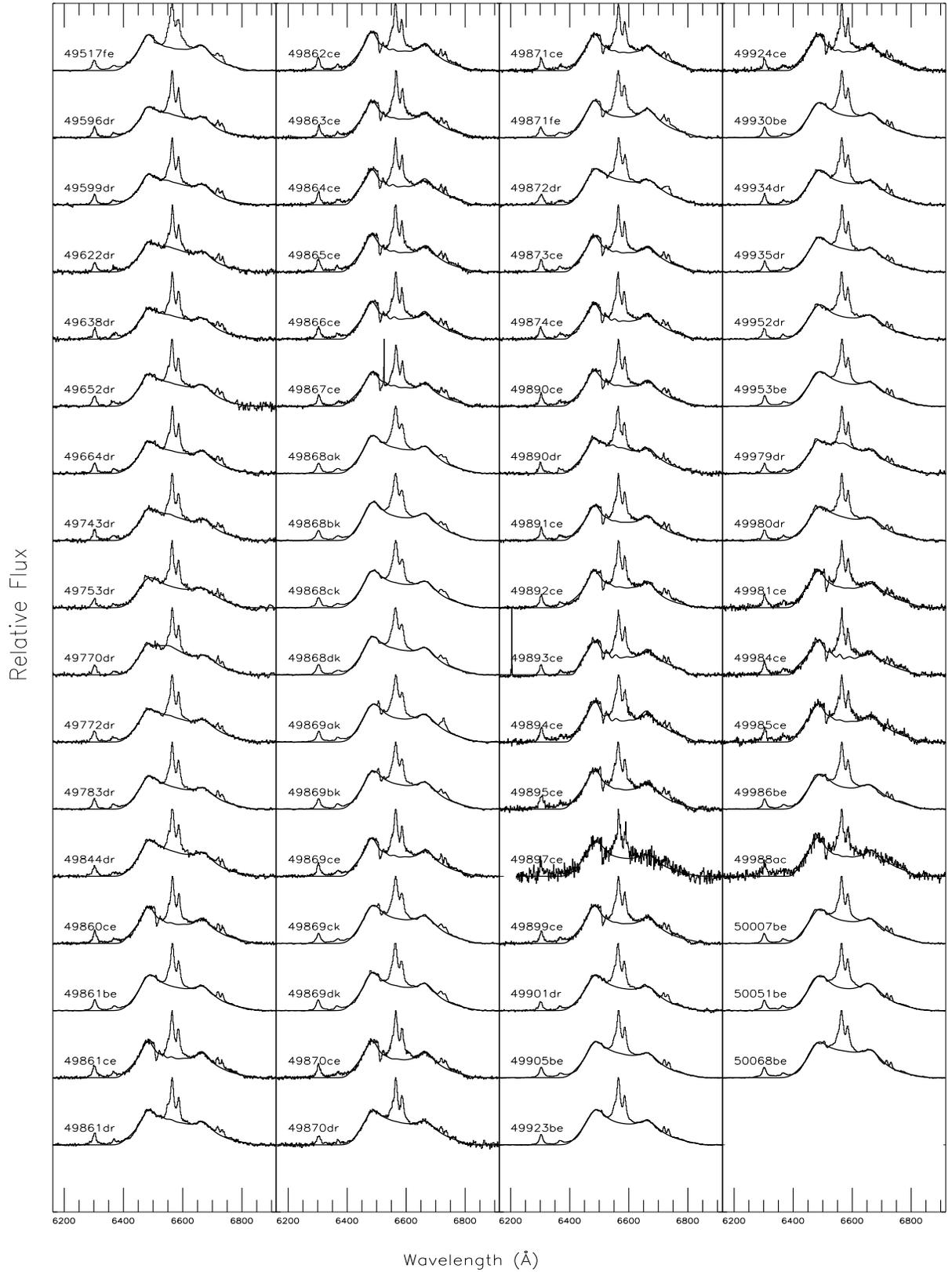}
\caption{The best fitted results for the double-peaked 
broad H$\alpha$ of 3C390.3. Thin solid line represents 
the observed spectrum, thick solid line represents the 
best fitted results for the double-peaked broad component 
of H$\alpha$. The corresponding ID is marked for each spectrum.}
\label{3c390}
\end{figure*}

\begin{figure*}
\centering\includegraphics[height = 22.5cm,width = 18cm]{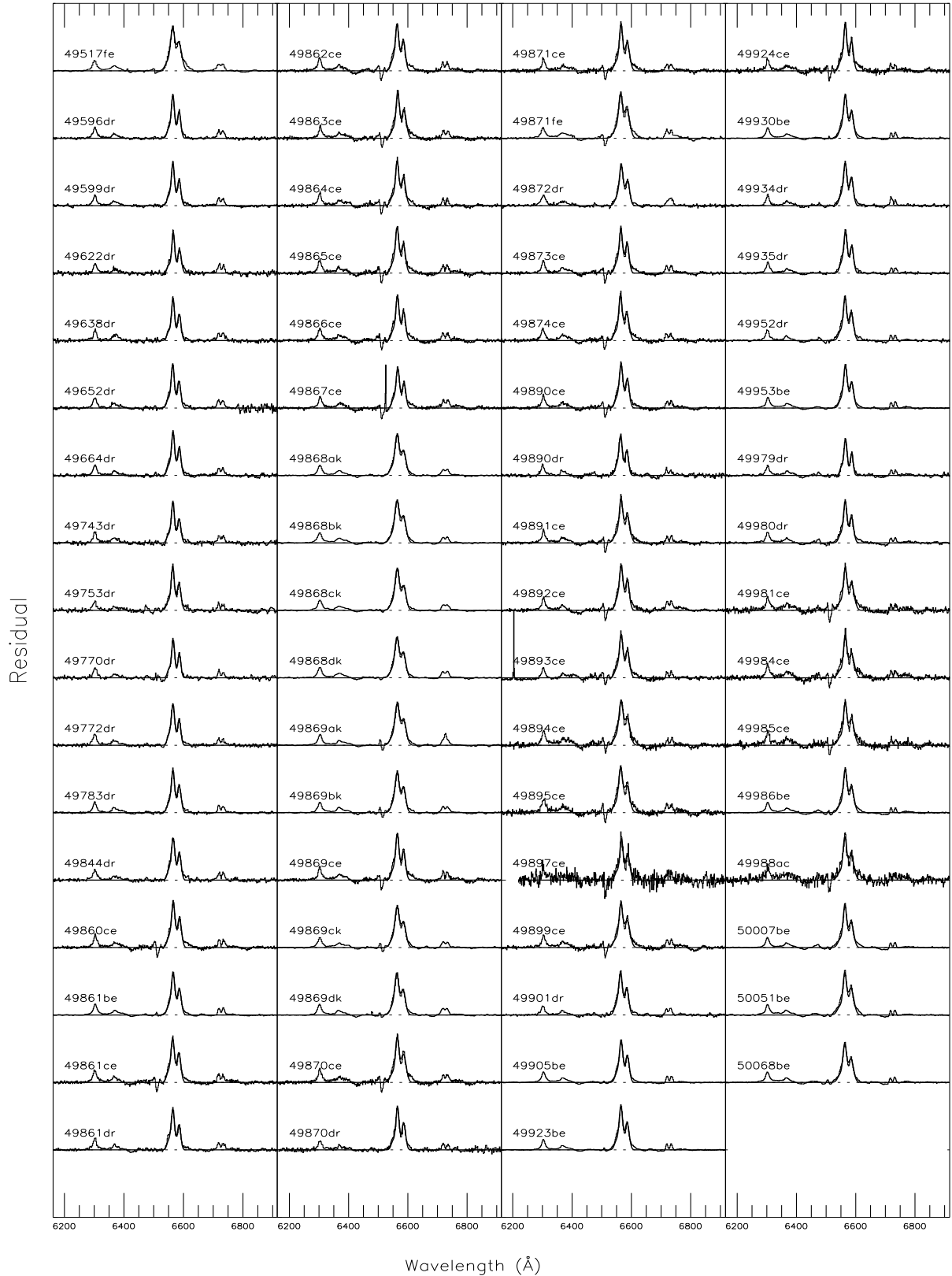}
\caption{The residuals (observed spectrum minus the broad 
double-peaked component of H$\alpha$) for the spectra 
shown in Figure~\ref{3c390}. Thin solid line represents 
the residual. Thick solid line represents the best 
fitted results for the narrow H$\alpha$ and 
[N~{\sc ii}]$\lambda6548,6583\AA$ doublet by standard gaussian 
functions. The corresponding ID is marked for each spectrum.}
\label{res}
\end{figure*}

    For 3C390.3, there are 67 public spectra observed around 
1995 including available H$\alpha$ which can be collected from 
the AGNWATCH project, based on those reported parameters listed in 
Table 9\ in Dietrich et al. (1998). Then the line parameters of 
the double-peaked broad H$\alpha$ are measured as follows. In order 
to ignore effects of the narrow lines, the double-peaked broad 
component of H$\alpha$ is firstly described and fitted by the 
theoretical elliptical accretion disk model proposed by 
Eracleous et al. (1995). Here, our main objective is not to 
accurately determine the disk parameters through theoretical model, 
thus, there are no further discussions about disk parameters, which 
has been discussed in detail in our another paper (Zhang 2011a, 
Paper I). In other words, we only require that the line profiles 
of the double-peaked broad H$\alpha$ can be best described and 
be well separated from the narrow lines. In this paper, we show 
the best fitted results for the apparent double-peaked broad 
H$\alpha$ in Figure~\ref{3c390}, and the corresponding residuals 
in Figure~\ref{res}. In Figure~\ref{res}, we also show the best 
fitted results for the narrow lines of H$\alpha$ and 
[N~{\sc ii}]$\lambda6548,6583$\AA\ by three standard gaussian 
functions with the same redshift and with line parameters 
ratio of [N~{\sc ii}] doublet fixed to theoretical values. Based 
on the results shown in Figure~\ref{3c390} and Figure~\ref{res}, 
the double-peaked broad component of H$\alpha$ can be clearly 
separated from the narrow lines.  
 
    Then, the line width (second moment) and the line flux of 
the double-peaked broad component can be estimated by 
(Peterson et al. 2004) 
\begin{equation}
\begin{split}
&\sigma_{H\alpha}^2  = \frac{\int_{\lambda}\lambda^2\times 
P_{\lambda}d\lambda}{\int_{\lambda}P_{\lambda}d\lambda} - 
(\frac{\int_{\lambda}\lambda\times P_{\lambda}d\lambda}
{\int_{\lambda}P_{\lambda}d\lambda})^2 \\
&flux(H\alpha) = \int_{\lambda}P_{\lambda}d\lambda
\end{split}
\end{equation}   
where $P_{\lambda}$ is the line profile of the double-peaked 
broad H$\alpha$ after subtraction of the narrow lines. Here, 
the main reasons to use the second moment rather than FWHM (full 
width at half maximum) are: the second moment rather than 
the FWHM can be used to well trace kepler velocities of broad 
line clouds in BLR of AGN (Fromerth \& Melia 2000, 
Peterson et al. 2004), the second moment rather than the FWHM is 
well defined for arbitrary line profiles (such as the double-peaked 
broad line profile), and moreover the second moment rather than 
the FWHM has relatively lower uncertainty (Peterson et al. 2004). 
The measured line widths ($\sigma$) of the double-peaked broad 
H$\alpha$ from the observed spectra are listed in Table 1. 

   Before proceeding further, it is necessary to estimate the 
corresponding uncertainty of the measured line width of the 
double-peaked broad H$\alpha$. Here, the commonly used bootstrap 
method (Peterson et al. 1998, 2004, Press et al. 1992, Zhang 2011b) 
is applied to estimate the uncertainty. Based on each observed 
double-peaked broad H$\alpha$ with narrow lines ([N~{\sc ii}], 
[S~{\sc ii}],[O~{\sc i}] doublets and narrow H$\alpha$) having 
been subtracted, 400 mock double-peaked lines can be created by 
the bootstrap method. When the bootstrap method is applied, we 
should avoid the selection of some adjacent data points to create 
mock lines, otherwise, the following estimated uncertainty should 
be unreasonably small, due to few variation of line profile. 
Then line widths of the 400 mock lines are measured by 
Equation (3). After that, standard deviation of the 400 measured 
line widths of the mock lines is accepted as the uncertainty 
of the line width of the observed double-peaked broad H$\alpha$, 
and listed in Table 1. Certainly, we should note that 
Peterson et al. (2004) accepted 
$\sigma_{err}/\sigma=81{\rm km/s}/3185{\rm km/s}\sim2.5\%$ 
as the mean uncertainty of the line width of broad H$\beta$ of 3C390.3 
within observed wavelength range from 5006\AA to 5246\AA. It is 
consistent with our value $<\sigma_{err}/\sigma>\sim2.2\%$, 
which indicates our estimated uncertainties for the line width 
of all the 67 double-peaked broad H$\alpha$ are reasonable and 
acceptable.
 
    Besides the line width and the corresponding uncertainty, 
it is not difficult to accomplish spectral flux calibration for 
3C390.3. However, because the spectra of 3C390.3 are observed by 
different instruments in different observatories under different 
configurations, the flux calibration based on the direct observed 
line profile should have low confidence level. As discussed in 
detail in the section 2 of Dietrich et al. (1998), some further 
procedures have been applied to complete the flux calibration. 
The spectra are firstly photometrically calibrated by comparison 
with the broadband photometric measurements (being convolved with 
the spectral response curve of the Johnson V and R), and then 
scaled by multiplicative factors to achieve the same total flux 
ratio as the intercalibrated photometric measurements in V and R 
bands. After that, the spectra have to be intercalibrated to a 
common flux level, in order to correct the effects for different 
wavelength shifts, and different spectral resolutions in different 
spectra. All the procedures have been successfully done by 
Dietrich et al. (1998), so that we directly collect the line 
fluxes ($f_1$) and corresponding uncertainties of broad H$\alpha$ 
listed in Dietrich et al. (1998). Moreover, in spite of the effects 
from different flux scales for the observed spectra, those directly 
measured line fluxes of the double-peaked broad H$\alpha$ ($f_0$, 
no corrections being considered) are also listed in Table 1.

     Before the end of the subsection, there are two points 
we should note. On the one hand, we do not try to confirm that 
the pure elliptical accretion disk model (Eracleous et al. 1995) 
is the unique true physical model for the double-peaked emitter 
3C390.3. We only use the theoretical disk model to separate the 
double-peaked broad component from the narrow lines. The properties 
of the residuals shown in Figure~\ref{res} for the best fitted 
results shown in Figure~\ref{3c390} illustrate that it is one 
good choice to determine the double-peaked broad component by 
the theoretical elliptical disk model for the spectra around 1995. 
On the other hand, although we do not apply the intercalibration 
method (Van Groningen \& Wanders 1992) to correct the effects from 
small wavelength shifts, different spectral resolutions and 
different flux scales for different spectra as what have been 
done in Dietrich et al. (1998), the measured second moment 
through observed spectrum is accurate and reliable, because 
the applied broadening velocity in the intercalibration 
method for 3C390.3 is only ten or more kilometers per second 
or smaller, which is can be totally ignored for the measured 
second moment and the corresponding uncertainty of the 
double-peaked broad H$\alpha$. Thus, the following results based 
on the measured second moments are reliable.

\subsection{On correlation between the line width and the line flux  
          of the double-peaked broad H$\alpha$}

\begin{figure*}
\centering\includegraphics[height = 9cm,width = 14cm]{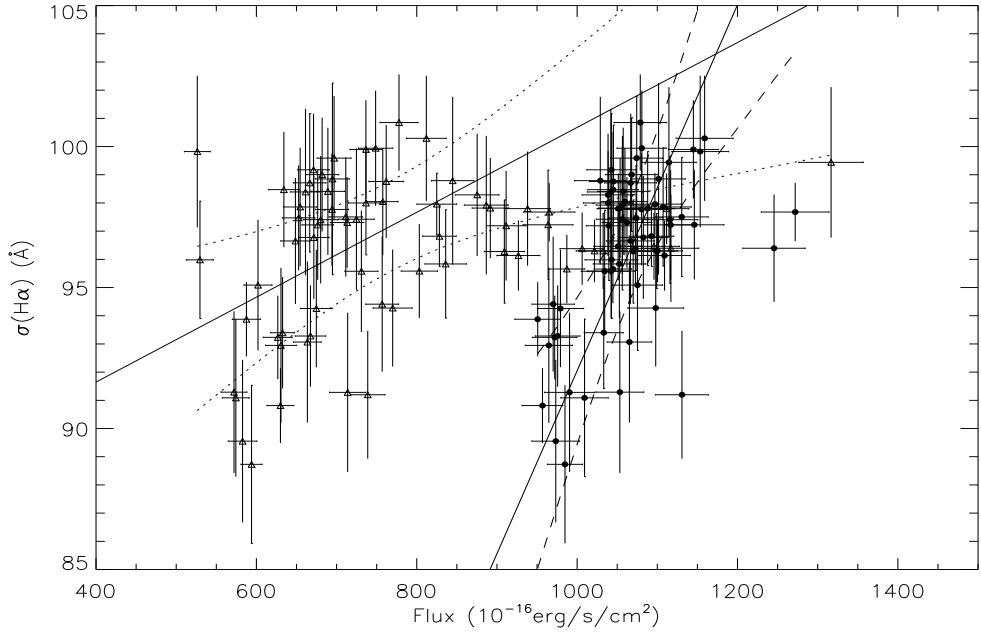}
\caption{The correlation between the line with (second moment) 
and the line flux of the double-peaked broad H$\alpha$ of 
3C390.3. Open triangles are for $f_0$ (no corrections for 
the different flux scales of different spectra) listed in 
table 1, solid circles are for $f_1$ collected from the 
AGNWATCH project.  Thin solid line and dotted lines represent 
the best fitted results for the open triangles and the 
corresponding 99.95\% confidence bands.  Thick solid line and 
dashed lines represent the best fitted results for the 
solid circles and the corresponding 99.95\% confidence bands. }
\label{fj_3c390}
\end{figure*}

   Figure~\ref{fj_3c390} shows the correlation between the line 
width and the line flux of the double-peaked broad H$\alpha$ of 
3C390.3. Linear correlation coefficient for the correlation of 
$f_0$ versus $\sigma$ is about 0.28 with two-sided significance 
of deviation from zero about $P_{null}\sim0.04$ for all the 67 
spectra listed in Table 1. The coefficient for the correlation of 
$f_1$ (accurate line flux of H$\alpha$ from the AGNWATCH project) 
versus $\sigma$ is about 0.45 with $P_{null}\sim0.0002$. It is clear 
that there is one  positive correction between the line width and 
the line flux of the double-peaked broad H$\alpha$ for 3C390.3. 
Furthermore, under considerations of the uncertainties of the parameters 
in both coordinates, the best fitted results for $f_1$ versus $\sigma$ 
can be written as 
\begin{equation}
\frac{\sigma}{{\rm \AA}}\propto (0.065\pm0.006)\times\frac{f_1}
{10^{-16}{\rm erg/s/cm^2}}
\end{equation}
Certainly, we should note the accurate flux $f_1$ collected from 
the AGNWATCH project includes the contributions from the narrow 
emission lines of $H\alpha$ and [N~{\sc ii}] doublet. However, the 
contributions do not affect the positive correlation, because the 
flux of the narrow emission lines of each spectrum is constant. The 
best fitted results are shown as solid lines in Figure~\ref{fj_3c390}. 
The corresponding 99.95\% confidence bands for the best fitted results 
are also shown in Figure~\ref{fj_3c390}. 

\begin{figure*}
\centering\includegraphics[height = 9cm,width = 14cm]{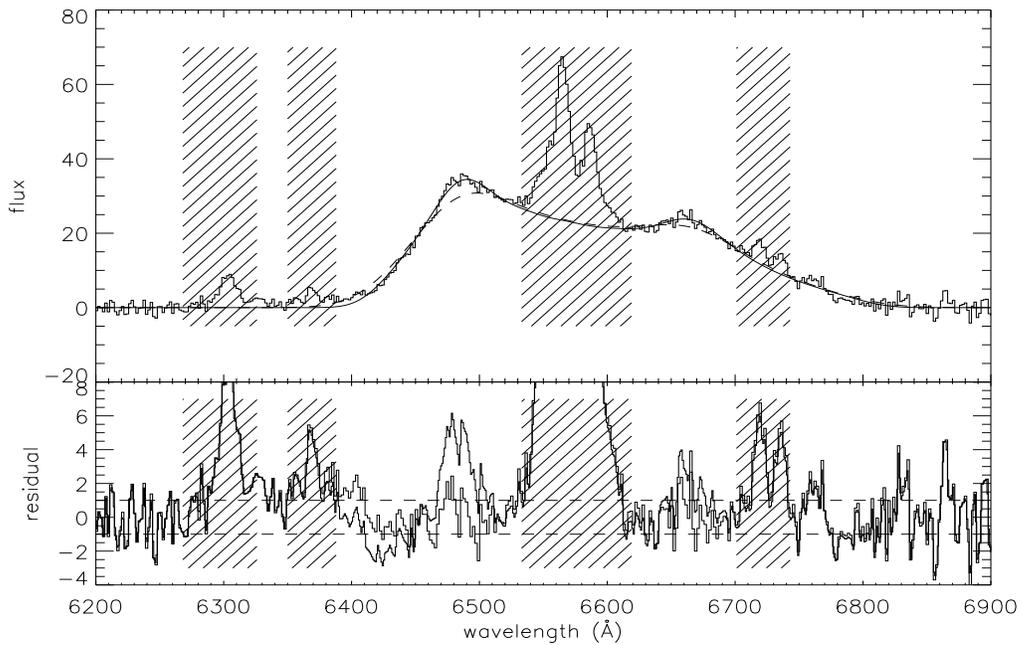}
\caption{In top panel, thin solid line represents the 
observed spectrum of '49870dr', thick solid line represents 
the best fitted results, thick dashed line represents the 
being broadened double-peaked broad line profile 
with $\sigma_{b}=1000{\rm km/s}$. In bottom panel, the 
corresponding residuals are shown, thin line is for the 
best fitted results, solid thick line is for the being 
broadened results. In the two panels, shadow areas are 
the masked regions for the narrow lines.}
\label{res2}
\end{figure*}

   %%%%%%%%effects of uncertainty on the coefficient%%%%%%
     It is clear that the uncertainty of the measured second 
moment ($\sigma_{err}$) of the double-peaked broad H$\alpha$ has 
important effects on our final conclusion about the positive 
correlation for 3C390.3, because the varying range of second moment 
is not large, $\sim90\AA<\sigma<\sim100\AA$. If the uncertainty 
of the second moment is large enough to cover the varying range of 
the second moment, the shown positive correlation in 
Figure~\ref{fj_3c390} should be fake. Thus, it is necessary to 
check whether larger uncertainties should be possible for our 
measured line width. Here, we simply consider the question as 
follows. For one given theoretical line profile of the double-peaked 
broad H$\alpha$ with second moment $\sigma$ (solid line shown in 
Figure~\ref{3c390}) best fitted for the observed line profile, if 
the corresponding uncertainty $\sigma_{err}$ for the second moment 
is reasonable, we will find that after being broadened by velocity 
$\sigma_{b}$ for the theoretical line profile, the new theoretical 
line profile with second moment $\sigma_{new}=\sigma+\sigma_{err}$ 
still well describes the observed double-peaked broad line profile 
of H$\alpha$. Here, the value of $\sigma_{b}$ can be determined 
by $\sigma_{b}^2 = \sigma_{new}^2 - \sigma^2$. Based on the 
listed values in Table 1, we can find that the common broadening 
velocity is around 20\AA\ ($\sim1000{\rm km/s}$). Figure~\ref{res2} 
shows one example about the being broadened line profile. In the 
figure, we can find that the being broadened line profile (the 
IDL function 'convol' being applied) with 
$\sigma_{b}=1000{\rm km/s}$ has begun to lead to some bad fitted 
results for the spectrum around the blue peak ($\sim6500$\AA) of 
the double-peaked broad H$\alpha$, which indicates larger 
broadening velocity should be unreasonable. In other words, 
the varying range of the second moment is larger enough to ignore 
the effects of the uncertainty of the second moment on the positive 
correlation shown in Figure~\ref{fj_3c390}, and the positive 
correlation is reliable.

\begin{figure*}
\centering\includegraphics[height = 9cm,width = 14cm]{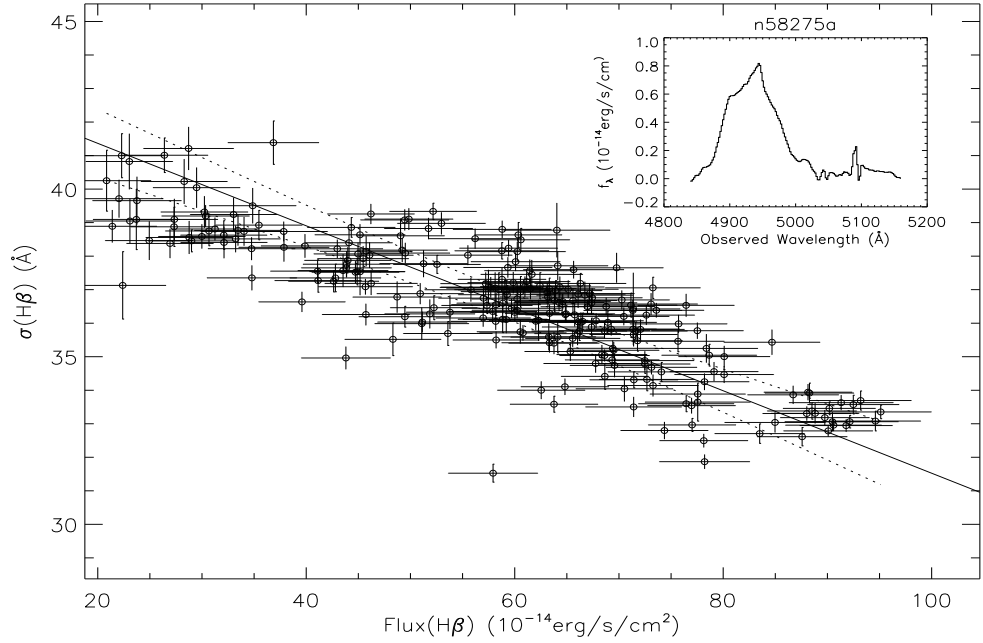}
\caption{The correlation between the line width and the 
line flux of the broad H$\beta$ of NGC5548. Solid line 
represents the best fitted results, dotted lines represent 
the corresponding 99.95\% confidence bands. In top-right panel, 
one line profile of the pure broad H$\beta$ is shown.}
\label{fj_ngc5548}
\end{figure*}

    Besides the unexpected positive correlation shown in 
Figure~\ref{fj_3c390} for 3C390.3, we wish to check the 
correlation of line parameters of broad lines for the other 
well-known mapped normal broad line AGN, in order to frankly 
show that the positive correlation for the double-peaked emitter 
3C390.3 is very interesting. There are more than 40 nearby mapped 
objects reported in the literature (Barth et al. 2011, 
Bentz et al. 2006, 2009, 2010, Denney et al. 2006, 2009, 2010, 
kaspi et al. 1996, 2000, 2005, 2007, Peterson et al. 2004). The 
simple correlation between the BLR size and the line width for 
part of those objects can be found in the corresponding literatures. 
Here, NGC5548 is selected as our target, because there is one 
well sample of fine spectra with homogeneous high-quality for 
NGC5548 (Bentz et al. 2007, Dietrich et al. 1993, Korista et al. 
1995, Peterson et al. 1991, 1992, 1994, 1999, 2002, 2004, 
Popovic et al. 2008, Sergeev et al. 2007, Shapovalova et al. 2004,  
Wanders \& Peterson 1996). Some detailed information about the 
spectra of NGC5548 can be found in the website of the AGNWATCH 
project. Here, we mainly focus on the spectra studied in detail 
in Wanders \& Peterson (1996) and collect the spectra with necessary 
corrections having been done by Wanders \& Peterson, i.e., 
the collected 247 spectra have been scaled through the van 
Groningen-Wanders method, and have the underlying continuum 
subtracted, and have the narrow components removed as well, and 
no further corrections are necessary for the following measured 
line parameters. Then based on the collected 247 spectra of 
NGC5548 observed from Dec. 1988 to Sep. 1993., the line width 
and the line flux can be easily determined by Equation (3), 
within observed wavelength range from 4840\AA\ to 5030\AA. 
Then, uncertainty of the line flux is determined by 
uncertainty in spectral flux provided by the AGNWATCH 
project, and uncertainty of the line width is determined by 
the bootstrap method. Here, we do not show all the observed 
spectra of NGC5548, but only show one spectrum as an example 
in the top-right panel of Figure~\ref{fj_ngc5548}. The measured 
line parameters and corresponding uncertainties of the broad 
H$\beta$ of NGC5548 are listed in Table 2. 

    Figure~\ref{fj_ngc5548} shows the correlation between the 
line width and the line flux of the broad H$\beta$ for NGC5548. 
The linear correlation coefficient is about -0.83 with 
$P_{null}\sim0$. The best fitted results with the considerations 
of the uncertainties of the parameters in both coordinates can be 
written as
\begin{equation}
\log(\frac{flux(H\beta)}{10^{-14}{\rm erg/s/cm^2}})\propto(-4.25\pm0.11)\times\log(\frac{\sigma(H\beta)}{{\rm \AA}})
\end{equation}
It is clear there is one strong negative correlation for NGC5548, 
which is well consistent with the expected result under the 
Virialization assumption: $L(H\beta_B)\propto\sigma(H\beta_B)^{-4}$. 
Undoubtfully, the negative correlation is well consistent with 
the negative correlation between the size of BLR and the line 
width found in Bentz et al. (2006), Denney et al. (2010) and 
Peterson et al. (2004).

    Based on the results above, we can find that there is one 
reliable positive correlation between the line width and the 
line flux of the double-peaked broad H$\alpha$ for the double-peaked 
emitter 3C390.3, which is against the expected negative correlation 
confirmed for the other mapped AGN. The unexpected positive 
correlation should indicate different structures of BLRs of the 
double-peaked emitter 3C390.3 and normal broad line AGN. 

\section{Discussions and Conclusions}

    It is clear that the reliable positive correlation shown 
in Figure~\ref{fj_3c390} for the double-peaked emitter 3C390.3 
is against the expected result through the Virialization 
assumption.  Thus the following three questions are mainly 
considered, what determines the unexpected positive correlation? 
whether the Virialization assumption is still available for 
double-peaked emitters? whether the unexpected positive 
correlation can be found for the other double-peaked emitters?

\subsection{Accretion Disk Model for Double-Peaked Emitters}

      For normal broad line AGN, such as NGC5548, the strong 
negative correlation of the broad line parameters can be naturally 
explained by the following simple viewpoint: BLR kinematics are 
dominated by central gravity (some more recent reviews can be 
found in  Down et al. 2010, Sluse et al. 2011), which could be 
well appropriate to the broad line Seyfert 1 galaxy NGC5548 
(Bentz et al. 2007, Denney et al. 2010, Peterson et al. 2002, 
2004, Wanders \& Peterson 1996,) and some other mapped broad 
line AGN (Peterson et al. 2004): stronger continuum emission 
leads to deeper ionization boundary (larger flux-weighted size 
of BLR) and leads to smaller general rotating velocity (smaller 
line width). And then, through the broad line strength being 
applied to estimate the size of BLR, 
$R_{BLR}\propto L(H\alpha)^{\sim0.5}$, the negative correlation 
of the line width and the line flux can be found. Certainly, 
the viewpoint is not appropriate to the double-peaked emitter 
3C390.3.

     If the theoretical accretion disk model was accepted, 
what result about the correlation between the size of BLR and 
the continuum/line flux should be expected for 3C390.3? In the 
context of the theoretical disk model for double-peaked emitters, 
such as the elliptical accretion disk model (Eracleous et al. 
1995) which can be well applied to best fit the observed line 
spectra of 3C390.3 observed around 1995, the disk-like BLR 
is locating into the central accretion disk, and does share 
the kinematic characters of the central accretion disk. For 
3C390.3, the inner and outer boundary radii of the elliptical 
disk-like BLR are about 
$R_{in}\sim150 - 250{\rm R_G}$ and $R_{out}\sim1100 - 1400{\rm R_G}$, 
under the theoretical accretion disk model (Eracleous \& Halpern 
1994, 2003, Flohic \& Eracleous 2008, Shapovalova et al. 2001, 
Zhang 2011a, Paper I). Based on the oversimplified extended 
disk-like BLR (no considerations of tiny contributions from 
probable hot spots, warped structures, spiral arms etc. 
for 3C390.3 around 1995), one simple result can be expected: 
stronger continuum emission leads to stronger broad line 
emission coming from the inner part of the disk-like BLR (and 
then larger line width) in one short period (much shorter 
than the traveling time for ionizing photons through the 
whole BLR), because the stronger ionizing continuum firstly 
reaches the inner part of the disk-like BLR. 
%{\bf  The expected result for 3C390.3 can be supported by the fact 
%that the profile wings respond to continuum variations on shorter time 
%scales than the profile core (Dietrich et al. 1998), although the estimated 
%time lag for line cores and line wings having large uncertainties.%}

\begin{figure*}
\centering\includegraphics[height = 9cm,width = 14cm]{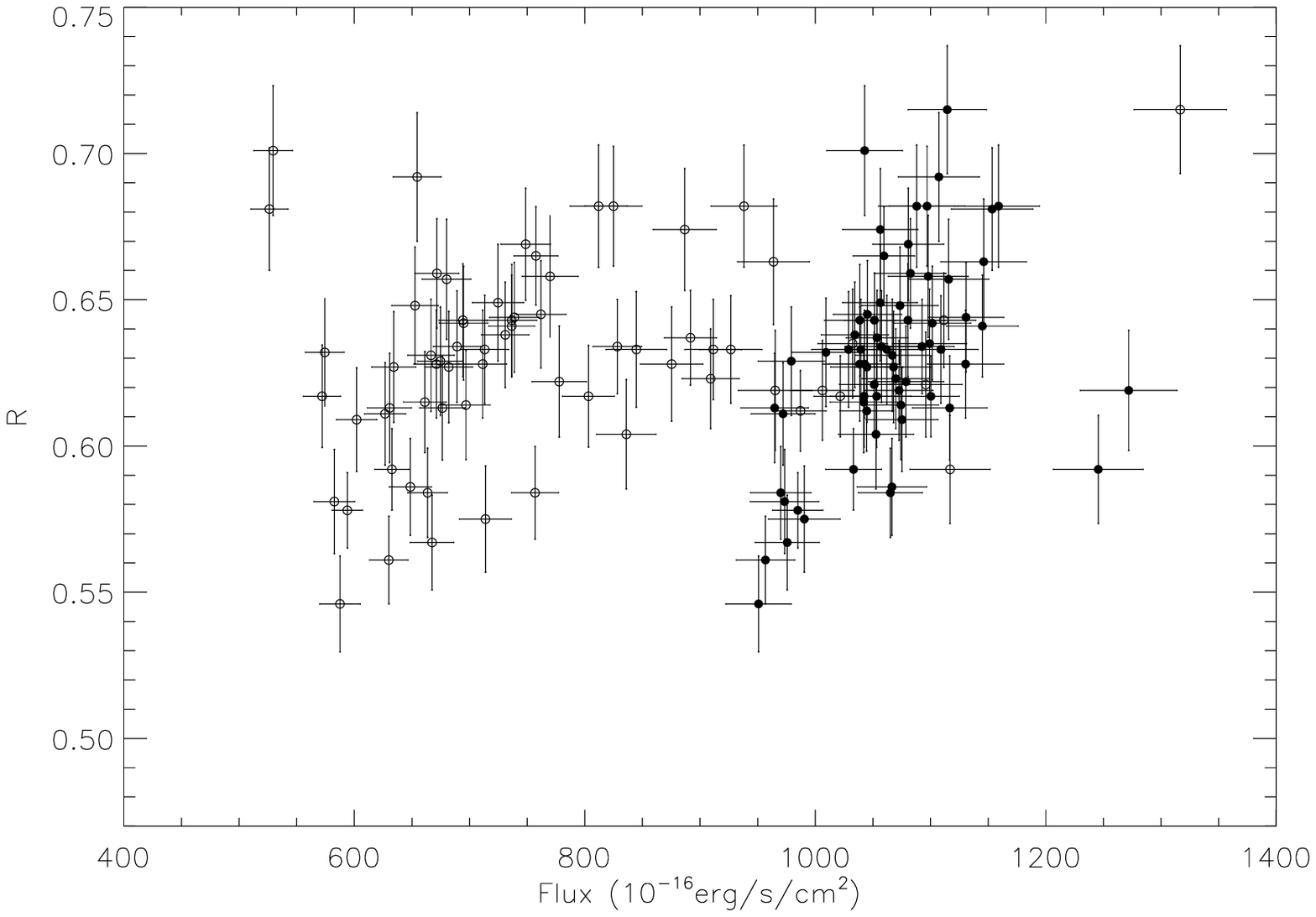}
\caption{The correlation between the line flux of 
the double-peaked broad H$\alpha$ and the parameter of $R$. 
Open circles are for $f_0$ in Table 1 and solid circles 
are for $f_1$ in Table 1.}
\label{fr}
\end{figure*}

   In other words, stronger continuum emission (strong line 
flux) leads to larger flux ratio of the broad H$\alpha$ from 
the inner part of the BLR to the broad H$\alpha$ from the 
outer part of the BLR, and leads to larger line width due 
to larger contributions from the higher velocity clouds in 
inner part of BLR, which is similar as the results shown in 
Figure~\ref{fj_3c390}. So that, we check the properties of 
the broad line flux ratio $R = f_{1/2}/f_{all}$, where $f_{1/2}$ 
means the broad line flux coming from the inner part of 
disk-like BLR with radius from $R_{in}$ to 
$0.5\times(R_{in} + R_{out})$ and $f_{all}$ means the measured 
total broad line flux. The calculated value of R for each 
spectrum is listed in Table 1. Figure~\ref{fr} shows the 
correlation between the line flux $f_0$ ($f_1$) and the 
parameter of $R$. In the figure, the shown uncertainty of 
$R$ is determined by the uncertainty of $f_1$ collected from 
the AGNWATCH project. The spearman-rank correlation coefficient 
for the correlation is about 0.49 with $P_{null}\sim10^{-5}$ 
for the parameter $f_1$ in Table 1 and 0.26 with 
$P_{null}\sim4\%$ for the parameter $f_0$ in Table 1. The 
positive correlation in Figure~\ref{fr} indicates the proposed 
disk origination for the double-peaked broad H$\alpha$ of 
3C390.3 can be naturally applied to explain the unexpected 
positive correlation shown in Figure~\ref{fj_3c390}.

   Moreover, if the observed broad H$\alpha$ of 3C390.3 was 
seperated into two parts: one part with larger line width 
($\sigma_1$) coming from the inner region of the disk-like BLR 
with radius from $R_{in}$ to $0.5\times (R_{in} + R_{out})$, 
and the other part with smaller line width ($\sigma_2$) coming 
from the outer region of the disk-like BLR, then the line 
width ($\sigma$) of the total observed double-peaked broad 
H$\alpha$ can be re-written as (Zhang 2011b)
\begin{equation}
\sigma^2 = R \times \sigma_1^2 + (1-R) \times \sigma_2^2 
\end{equation}
where $R$ is the flux ratio of $R = f_{1/2}/f_{all}$. The 
equation above can be easily proved by definition of second 
moment (Equation 3). Based on the mean disk parameters shown 
in our paper I (Zhang 2011a), we have 
$\sigma_1\sim109{\rm \AA}$ and $\sigma_2\sim69{\rm \AA}$. Through 
the equation above, in order to obtain the measured line width 
of the double-peaked broad H$\alpha$ varying from $\sim92$\AA\ to 
$\sim100$\AA (minimum and maximum values for the second 
moment of broad H$\alpha$ of 3C390.3), the parameter of $R$ 
should be varying from $\sim0.53$ to $\sim0.73$, which is 
consistent with the shown results about R in Figure~\ref{fr} 
and in Table 1.

   The results above indicate that the unexpected positive 
correlation shown in Figure~\ref{fj_3c390} for 3C390.3 can 
be naturally explained by the proposed accretion disk 
origination of the double-peaked broad H$\alpha$. Certainly, 
here we do not consider effects from probable 'external' 
ionizing source for the double-peaked emitter 3C390.3, which 
will be discussed in detail in subsection 3.4.

\subsection{Virialization Assumption}

    As the most convenient method to estimate central black 
hole masses of broad line AGN, the virialization assumption 
has been widely and commonly accepted. Here, we should find 
that the assumption is also valid for the double-peaked 
emitter 3C390.3. 

    As we have shown above, the proposed accretion disk model 
should be appropriate for the double-peaked emitter 3C390.3.  
Considering the oversimplified case, the disk-like BLR of 3C390.3 
is separated into one inner region with radius from $R_{in}$ to 
$0.5\times (R_{in} + R_{out})$ and the other outer region, then 
the flux weighted distance between the two regions to the central 
black hole can be calculated based on the simple elliptical 
accretion disk model (Eracleous et al. 1995) 
$R_{BLR,1}\sim445{\rm R_G}$ for the inner part of the BLR and 
$R_{BLR,2}\sim987{\rm R_G}$ for the outer part of the BLR. 
According to the results shown in Zhang (2011b), the size of 
total BLR ($R_{BLR}$) can be written as
\begin{equation}
R_{BLR} = R \times R_{BLR,1} + (1-R)\times R_{BLR,2}
\end{equation}
Based on the values of R listed in Table 1 and the values of 
$R_{BLR,1}$ and $R_{BLR,2}$, we will have the size of BLR 
$R_{BLR}$ varying from $\sim596{\rm R_G}$ to $\sim715{\rm R_G}$. 
Moreover, based on the Virialization assumption, 
$R_{BLR}\times\sigma^2 = constant$ (or 
$(\frac{\sigma_{max}}{\sigma_{min}})^2=\frac{R_{BLR,max}}{R_{BLR,min}}$), 
combining with the measured line width in Table 1, the expected 
maximum variation of the size of BLR of 3C390.3 calculated through 
the variations of the line width should be 
$(\frac{\sigma_{max}}{\sigma_{min}})^2=(\frac{\sim100{\rm \AA}}{\sim92{\rm \AA}})^2\sim1.18$, 
which is consistent with the results above 
$\frac{R_{BLR,max}}{R_{BLR,min}}\frac{\sim715{\rm R_G}}{\sim596{\rm R_G}}\sim1.19$. 

    The results above indicate that the virialization assumption 
is still available for the double-peaked emitter 3C390.3. However, 
the common empirical relation $R_{BLR}\propto L^{\sim0.5}$ is not 
available to show the properties of {\bf time-varying} size of 
BLRs of 3C390.3. In other words, based on the {\bf mean} continuum 
luminosity (broad line luminosity) in one long period, the {\bf mean} 
size of BLR can be estimated by the empirical relation 
$R_{BLR}\propto L^{\sim0.5}$, however, the time-varying size of 
BLR is decreasing with the line luminosity increasing.

\subsection{Previous Results}

   Besides the well-known double-peaked emitter 3C390.3, 
there are several other double-peaked emitters, of which 
the double-peaked broad line profiles have been studied 
and reported. Some simple results about correlation between 
peak separation (in some literature, peak separation was 
used as substitute for line width) and line flux (continuum 
emission) can be found. 

    In Figure 4 of Shapovalova et al. (2001), we can find 
one rough negative correlation between peak separation and 
line flux (continuum emission) for the double-peaked broad 
H$\beta$ of 3C390.3 observed from 1995 to 2000. Similarly 
negative correlation can be found for the double-peaked 
broad H$\alpha$ of NGC1097 (Schimoia et al. 2012, 
Storchi-Bergmann et al. 2003). However, through study of one 
sample of double-peaked emitters, Lewis et al. (2010) have 
shown that there is no confirmed correlation between the peak 
separation and the line flux for long-term observed double-peaked 
broad H$\alpha$, and additionally, the expected negative 
correlation between the line width (either FWHM or FWQM) and 
the broad H$\alpha$ flux is not observed. The contradictory 
is mainly due to the following reason: there are important 
effects on peak separation from varying of physical disk 
parameters, such as the effects from accretion disk precession, 
in the context of the accretion disk model for double-peaked 
emitters.

   As we have seen that the proposed accretion disk models 
for double-peaked emitters lead to one important result: 
disk precession (or physical disk structure parameters varying) 
dominates the line profile variations of double-peaked broad line, 
besides the not so important contributions from the other 
subtle structures, such as the hot spots, warped structures etc.. 
In some cases, such as disk-like BLR with large eccentricity 
$e\sim0.4$ under the elliptical accretion disk model, even 
there were no variations of continuum emission, only the 
orientation angle of elliptical rings $\phi_0$ varying 
(pure accretion disk precession) should lead line width 
(second moment) varying by 30\%, which can be simply verified 
by the proposed theoretical models. Thus, when we consider the 
correlation of the line parameters of double-peaked broad lines, 
the effects from the accretion disk precession (or from the 
other physical disk parameters varying) should be firstly 
corrected. However, before confirming one clear theoretical 
model for one double-peaked emitter, it is hard to correct 
these effects. One good choice to ignore the effects of accretion 
disk parameters varying is to consider short-term observed 
spectra with few variations of line profiles. That is the 
main reason why only the spectra around 1995 are collected 
for the double-peaked emitter 3C390.3, because the line profiles 
around 1995 are almost stable with less effects from disk 
parameters varying, as what we have discussed in our Paper I 
(Zhang 2001a). However,  in Lewis et al. (2010), Schimoia et al. 
(2012), Shapovalova et al. (2001) and Storchi-Bergmann et al. 
(2003), the observed double-peaked broad lines have apparent 
variations of both peak flux ratios and peak positions, 
which indicate apparent variations of accretion disk physical 
parameters. If only continuum variations are considered, 
there should be much weak variations for peak positions and 
peak flux ratios, besides apparent variations of line intensity.  

    Moreover, We further check the listed parameters in the 
literature for the double-peaked emitter 3C390.3, under the 
virialization assumption: $R_{BLR}\times V^2(BLR)=constant$. 
In Wandel et al. (1999) and Peterson et al. (2004), the size 
of BLR and the corresponding line width are 
$R_{BLR,2004}\sim24\pm6{\rm light-days}$ and 
$FWHM(2004)\sim10000{\rm km/s}$ for 3C390.3 around 1995. 
In Shapovalova et al. (2001), the size of BLR and the 
corresponding line width are 
$R_{BLR,2001}\sim90\pm10{\rm light-days}$ (similar results 
can be found in Popovic et al. 2011, some larger results 
can be found in Sergeev et al. 2011) and 
$FWHM(2001)\sim12000{\rm km/s}$ for 3C390.3 from 1995 to 
2000. It is clear that based on the measured line width 
(FWHM) and the size of BLR, we have 
$\frac{R_{BLR,2001}}{R_{BLR,2004}}\sim 3.8 >> 
(\frac{FWHM(2004)}{FWHM(2001)})^2 \sim 0.5$, which indicates 
the measured line width (FWHM) varying does not obey the 
virialization assumption. In other words, in one long-term 
observed period, the effects of accretion disk physical 
parameter varying should be important, especially on the 
line profile varying, so that the measured line width 
(FWHM) can not accurately trace the kepler velocity of broad 
line clouds for long-term spectra. 

   In order to more clearly show our point: the positive 
correlation between the line width and the line flux is 
not fake, some evidence to the contrary is shown as follows. 
If we accept that the positive correlation shown in 
Figure~\ref{fj_3c390} was fake for 3C390.3 and the common 
negative correlation could be applied for the double-peaked 
emitters, i.e., both the empirical relation 
$L^{\sim0.5}\propto R_{BLR}$ for time-varying size of BLR 
and applying peak separation ($V_p$) to replace the second 
moment were available for 3C390.3, we will have 
$L\times V_p^4 = constant$, under the virialization assumption. 
For the spectra of 3C390.3\ in Shapovalova et al. (2001), the 
line luminosity ratio of maximum $L(H\beta)$ to minimum 
$L(H\beta)$ is about $\frac{L(max)}{L(min)}\sim1.75$, 
however, the forth power of the line width ratio of minimum 
$V_p$ to maximum $V_p$ is about 
$(\frac{V_p(min)}{V_p(max)})^4\sim(\frac{8000{\rm km/s}}{14000{\rm km/s}})^4\sim0.11$, 
which is against the expected result 
$\frac{L(max)}{L(min)}=(\frac{V_p(min)}{V_p(max)})^4$.  
Similarly, for NGC1097 in Schimoia et al. (2012) and  
and Storchi-Bergmann et al. (2003), the line luminosity 
ratio of maximum $L(\alpha)$ to minimum 
$L(H\alpha)$ is about $\frac{L(max)}{L(min)}\sim3.7$, 
however, the forth power of the line width ratio of 
minimum $V_p$ to maximum $V_p$ is about 
$(\frac{V_p(min)}{V_p(max)})^4\sim(\frac{6000{\rm km/s}}{10000{\rm km/s}})^4\sim0.13$. 
Similar results can be found for the double-peaked emitters 
discussed in Lewis et al. (2010): the parameter 
$(\frac{V_p(min)}{V_p(max)})^4$ is much different from 
the parameter $\frac{L(max)}{L(min)}$, 
$(\frac{V_p(min)}{V_p(max)})^4\sim0.16$ and 
$\frac{L(max)}{L(min)}\sim1.4$ for 3C59, 
$(\frac{V_p(min)}{V_p(max)})^4\sim0.06$ and 
$\frac{L(max)}{L(min)}\sim2.2$ for IE0450.3,
$(\frac{V_p(min)}{V_p(max)})^4\sim0.15$ and 
$\frac{L(max)}{L(min)}\sim2.5$ for Pictor A,
$(\frac{V_p(min)}{V_p(max)})^4\sim0.18$ and 
$\frac{L(max)}{L(min)}\sim1.7$ for CBS74,
$(\frac{V_p(min)}{V_p(max)})^4\sim0.31$ and 
$\frac{L(max)}{L(min)}\sim2.1$ for PKS0921,
$(\frac{V_p(min)}{V_p(max)})^4\sim0.53$ and 
$\frac{L(max)}{L(min)}\sim1.8$ for PKS1020,
$(\frac{V_p(min)}{V_p(max)})^4\sim0.27$ and 
$\frac{L(max)}{L(min)}\sim2.5$ for PKS1739.  
The results re-confirm that for long-term observed double-peaked 
broad lines, the parameter of the peak separation is not one 
good substitute to trace the kepler velocity of the broad 
line clouds, and clearly indicates the correlation shown 
in Figure~\ref{fj_3c390} is definitely different from the 
previous reported results about the peak separation in 
the literature.

   The results above indicate that for double-peaked 
emitters, it should be one good choice to check the 
virialization assumption (the correlation of line parameters) 
by one sample of short-term observed spectra rather than by 
one sample of long-term observed spectra, in order to ignore 
the effects of physical disk parameters varying. 

\subsection{Indicator for Double-Peaked Emitters}

   Finally, we consider the question: whether the positive 
correlation of line parameters could be common for all 
double-peaked emitters, if the effects of BLR physical 
parameter varying could be clearly considered?

   The most important starting point for the positive correlation 
is that the central ionizing continuum emission is coming from 
one point-like source, so that there are different arriving times 
of ionizing photos in  different parts of disk-like BLRs of 
double-peaked emitters. However, as discussed in the literature, 
one 'external' ionizing source should be necessary for some 
double-peaked emitters (Chen et al. 1989, Eracleous \& Halpern 1994, 
2003, Luo et al. 2009, Strateva et al. 2003, 2006, 2008, Wu et al. 2008), 
due to the 'energy budget' problem: the total double-peaked broad line 
flux likely exceeds the viscous energy that can be extracted locally 
from the accretion disk. If the extended size of the 'external' 
ionizing source is large enough to cover the disk-like BLRs 
of double-peaked emitters, the expected different time delays 
for different parts of BLRs should be not so apparent, and then 
no apparent positive correlation between the line width and the 
line flux of broad double-peaked lines could be found. 

   Commonly, parameter $L(H\alpha)/W_d>0.2$ (ratio of the 
double-peaked broad H$\alpha$ luminosity to the viscous power 
output of the line-emitting portion of the accretion disk) is 
used to determine existence of one 'external' ionizing source 
(Collin-Souffrin 1987, Dumont \& Collin-Souffrin 1990, 
Strateva et al. 2008, Williams 1980). For 3C390.3, Eracleous \& Halpern 
(2003, 1994) have shown that the parameter $L(H\alpha)/W_d$ is about  
$3.5\times10^{42} {\rm erg/s}/1.6\times10^{43} {\rm erg/s}\sim0.22$  
with energy transfer efficiency $\eta\sim0.01$ being accepted. 
Here the used $L(H\alpha)$ and $W_d$ are the updated values 
listed in Eracleous \& Halpern (2003).  However,  we should 
note that the energy transfer efficiency is much larger than 
0.01 for 3C390.3. Based on the empirical relation shown in 
Davis \& Laor (2011) and the more recent  black hole mass of 
3C390.3\ in Dietrich et al. (2012, $M_{BH}\sim10^9{\rm M_{\odot}}$), 
the energy transfer efficiency for 3C390.3 is about 
$\eta\sim0.089\times(M_{BH,8})^{0.52}\sim0.29$. So 
that the parameter of $L(H\alpha)/W_d$ is about 
$L(H\alpha)/W_d\sim0.0075$, which is much smaller than the 
limited value of $L(H\alpha)/W_d\sim0.2$.  So that, it is 
not necessary for one 'external' ionizing source for 
the double-peaked emitter 3C390.3. Furthermore, through 
the reverberation mapping technique (Blandford \& Mckee 1982, 
Peterson 1993), the UV emission region is within 5 light-days 
(Dietrich et al. 1998), which is much smaller than the size 
of BLR ($R_{BLR}\sim20 {\rm light-days}$ in Peterson et al. (2004) 
(certainly, much smaller than the size of BLR shown in 
Dietrich et al. 2012, Popovic et al. 2011, 
Sergeev et al. 2011 and Shapovalova et al. 2001). So that the 
time delays for different parts of the BLR of 3C390.3 are 
apparent, and one positive correlation shown in 
Figure~\ref{fj_3c390} can be expected.

    So far, It is still not clear about the structures of 
the probable needed 'external' ionizing source: external 
ionizing photos coming from an elevated structure in central 
region of accretion disk (standard RIAF, Cao 2005, Manmoto 2000), 
or ionizing photos coming from scattering ionizing photons 
radiated from the inner accretion disk region (Cao \& Wang 2006), 
or other unknown structures. Before to confirm the structures of 
the 'external' ionizing source, it is very hard to give one 
final conclusion that all double-peaked emitters show one 
positive correlation between line width and line flux for 
short-term observed spectra. However, after the corrections 
of the effects of disk physical parameters varying, once one 
positive correlation between the line width and the line flux 
can be confirmed for one broad line, it is extremely possible 
that the broad line comes from central accretion disk.

\subsection{Conclusions}

The final conclusions are as follows. 
\begin{itemize}
\item Under the Virialization assumption and the commonly 
accepted empirical relation $R_{BLR}\propto L^{\sim0.5}$, 
there should be one strong negative correlation between the 
line width and the line flux of broad lines for 
multi-observed spectra of broad line AGN: 
$L\times \sigma^4 = constant$. The negative correlation can 
be confirmed for some mapped objects, such as the results 
shown in Figure~\ref{fj_ngc5548} for NGC5548.
\item Based on the public spectra of 3C390.3 from the 
AGNWATCH project, the correlation between the line width 
(second moment) and the line flux of the double-peaked broad 
H$\alpha$ is checked. However, one unexpected reliable positive 
correlation is confirmed and shown in Figure~\ref{fj_3c390}, 
which indicates there are different structures of BLR for the 
double-peaked emitter 3C390.3 and normal broad line AGN. 
\item In the context of the theoretical accretion disk model 
for double-peaked emitters, the unexpected positive correlation 
between the line width and the line flux can be naturally 
explained, due to the different time delays for the different 
parts of the disk-like BLRs of 3C390.3\ in one short period. 
\item The Virialization assumption is further checked for 
the double-peaked emitter 3C390.3, 
$R_{BLR}\times\sigma^2=constant$. Then, we find that 
the virialization assumption is still available for 3C390.3, 
however, the time-varying size of BLR in one period can 
not be expected through the empirical relation of 
$R_{BLR}\propto L^{\sim0.5}$. For the double-peaked 
emitter 3C390.3, stronger continuum emission leads to 
smaller (not larger) flux-weighted size of BLR in one moment.
\item Furthermore, we have shown that in the context of 
the accretion disk model, there are important effects on 
the correlation of the line parameters from the physical 
disk parameters varying (such as the accretion disk precession). 
That is why there are different conclusions in the 
literature about the correlation between the peak separation 
and the line flux for long-term observed double-peaked 
broad lines. Moreover, it should be better to check the 
correlation of line parameters of double-peaked broad 
line through short-term observed spectra, in order 
to ignore the effects from physical disk parameters varying. 
\item Finally, the correlation between the line width and 
the line flux is checked for the other double-peaked emitters. 
However, due to the probable 'external' ionizing source 
with so far unclear structures, it is hard to confirm that 
the positive correlation between the line width and 
the line flux of double-peaked broad lines can be found 
for all the double-peaked emitters. However, one strong positive 
correlation between the line width and the line flux of one 
broad line should strongly indicate accretion disk 
origination for the broad line.  
\end{itemize}    

\section*{Acknowledgments}
Zhang X. G. gratefully acknowledge the anonymous 
referee for giving us constructive comments and 
suggestions to greatly improve our paper. ZXG gratefully 
acknowledges the finance support from the Chinese grant 
NSFC-11003043, and thanks the project of AGNWATCH
(http://www.astronomy.ohio-state.edu/\~{}agnwatch/) to make us
conveniently collect the spectra of 3C390.3 and NGC5548.

\clearpage
%\begin{landscape}
\begin{table}
\centering
\begin{minipage}{170mm}
\caption{Line Parameters of the double-peaked broad H$\alpha$ of 3C390.3}
\begin{tabular}{lllll|lllll}
\hline
id & $f_0$  & $f_1$ & $\sigma$  & R & id & $f_0$  & $f_1$ & $\sigma$ & $R$ \\
\hline
49517fe  &  632.6$\pm$14.8  &  1033.1$\pm$24.2  &  93.18$\pm$1.97  &  0.592  &  
49871ce  &  680.0$\pm$21.2  &  1115.7$\pm$34.8  &  97.31$\pm$2.25  &  0.657  \\
49596dr  &  587.5$\pm$17.5  &  950.7$\pm$28.4  &  93.44$\pm$1.32  &  0.546 & 
49871fe  &  736.5$\pm$19.9  &  1144.9$\pm$31.0  &  99.63$\pm$1.74  &  0.641\\
49599dr  &  629.9$\pm$16.8  &  956.7$\pm$25.5  &  91.56$\pm$1.33  &  0.561  &  
49872dr  &  671.6$\pm$19.0  &  1082.5$\pm$30.7  &  95.81$\pm$2.18  &  0.659  \\
49622dr  &  713.7$\pm$22.5  &  990.5$\pm$31.2  &  91.06$\pm$2.82  &  0.575  &
49873ce  &  694.2$\pm$20.8  &  1080.3$\pm$32.3  &  97.55$\pm$1.98  &  0.643\\
49638dr  &  756.8$\pm$20.5  &  970.0$\pm$26.2  &  94.11$\pm$2.39  &  0.584  & 
49874ce  &  724.7$\pm$22.2  &  1056.2$\pm$32.4  &  97.76$\pm$2.51  &  0.649\\
49652dr  &  593.9$\pm$13.2  &  984.8$\pm$21.8  &  89.97$\pm$2.81  &  0.578  &
49890ce  &  676.4$\pm$19.6  &  1116.6$\pm$32.4  &  97.92$\pm$1.95  &  0.613\\
49664dr  &  667.4$\pm$19.0  &  975.6$\pm$27.8  &  93.51$\pm$1.79  &  0.567  & 
49890dr  &  694.8$\pm$21.0  &  1101.5$\pm$33.3  &  95.18$\pm$3.37  &  0.642 \\
49743dr  &  630.6$\pm$19.2  &  964.7$\pm$29.4  &  92.67$\pm$2.75  &  0.613  &
49891ce  &  689.1$\pm$20.6  &  1057.4$\pm$31.7  &  98.44$\pm$2.26  &  0.634\\
49753dr  &  574.3$\pm$16.8  &  1009.2$\pm$29.6  &  91.39$\pm$2.79  &  0.632  & 
49892ce  &  648.6$\pm$18.3  &  1066.5$\pm$30.1  &  96.89$\pm$2.23  &  0.586\\
49770dr  &  582.6$\pm$17.8  &  973.3$\pm$29.7  &  89.46$\pm$2.88  &  0.581  &
49893ce  &  529.6$\pm$16.7  &  1042.8$\pm$32.9  &  95.98$\pm$2.09  &  0.701 \\
49772dr  &  674.7$\pm$19.8  &  979.1$\pm$28.8  &  94.10$\pm$2.08  &  0.629  &
49894ce  &  938.1$\pm$28.7  &  1087.9$\pm$33.2  &  98.68$\pm$2.06  &  0.682\\
49783dr  &  626.6$\pm$17.9  &  972.0$\pm$27.8  &  93.77$\pm$1.48  &  0.611  &
49895ce  &  844.7$\pm$26.3  &  1028.9$\pm$32.1  &  98.21$\pm$2.95  &  0.633\\
49844dr  &  663.5$\pm$17.3  &  1065.1$\pm$27.7  &  93.47$\pm$2.85  &  0.584  &
49897ce  &  572.0$\pm$16.2  &  1053.2$\pm$29.8  &  96.29$\pm$2.87  &  0.617\\
49860ce  &  681.9$\pm$20.6  &  1067.9$\pm$32.3  &  98.31$\pm$2.02  &  0.627  &
49899ce  &  777.8$\pm$23.7  &  1078.7$\pm$32.8  &  100.44$\pm$1.71  &  0.622\\
49861be  &  836.0$\pm$25.8  &  1052.6$\pm$32.5  &  95.91$\pm$1.92  &  0.604  & 
49901dr  &  803.1$\pm$22.6  &  1042.3$\pm$29.3  &  94.93$\pm$1.66  &  0.617\\
49861ce  &  696.9$\pm$21.1  &  1074.3$\pm$32.6  &  98.06$\pm$2.21  &  0.614  &
49905be  &  909.2$\pm$24.8  &  1069.8$\pm$29.2  &  95.73$\pm$1.83  &  0.623\\
49861dr  &  602.0$\pm$17.5  &  1075.2$\pm$31.2  &  95.24$\pm$2.32  &  0.609  &
49923be  &  891.6$\pm$22.6  &  1053.5$\pm$26.7  &  97.21$\pm$1.78  &  0.637\\
49862ce  &  634.2$\pm$19.1  &  1044.7$\pm$31.5  &  97.87$\pm$2.04  &  0.627  &
49924ce  &  748.6$\pm$21.4  &  1080.6$\pm$30.9  &  99.44$\pm$2.04  &  0.669\\
49863ce  &  666.5$\pm$20.2  &  1066.8$\pm$32.3  &  98.45$\pm$2.46  &  0.631  &
49930be  &  911.4$\pm$24.6  &  1039.5$\pm$28.1  &  96.89$\pm$1.94  &  0.633\\
49864ce  &  886.7$\pm$27.3  &  1056.4$\pm$32.6  &  97.67$\pm$2.44  &  0.674  &
49934dr  &  736.6$\pm$21.8  &  1038.6$\pm$30.7  &  96.31$\pm$1.84  &  0.643\\
49865ce  &  875.3$\pm$27.2  &  1038.5$\pm$32.2  &  98.19$\pm$2.16  &  0.628  &
49935dr  &  713.1$\pm$20.8  &  1062.1$\pm$31.0  &  96.79$\pm$2.41  &  0.633\\
49866ce  &  761.9$\pm$21.6  &  1045.1$\pm$29.7  &  98.05$\pm$1.99  &  0.645  &
49952dr  &  652.6$\pm$20.1  &  1073.4$\pm$33.1  &  96.66$\pm$1.64  &  0.648\\
49867ce  &  730.9$\pm$20.6  &  1034.3$\pm$29.1  &  95.54$\pm$2.15  &  0.638  &  
49953be  &  757.5$\pm$19.1  &  1059.5$\pm$26.7  &  96.94$\pm$1.91  &  0.665\\
49868ak  &  1006.2$\pm$27.6  &  1072.6$\pm$29.5  &  96.08$\pm$1.28  &  0.619  &
49979dr  &  769.9$\pm$24.3  &  1097.9$\pm$34.7  &  93.91$\pm$2.08  &  0.658\\
49868bk  &  1032.4$\pm$30.2  &  1099.1$\pm$32.1  &  95.21$\pm$1.33  &  0.635  & 
49980dr  &  711.4$\pm$20.8  &  1130.4$\pm$33.1  &  95.04$\pm$2.11  &  0.628\\
49868ck  &  1095.8$\pm$31.6  &  1051.1$\pm$30.3  &  95.85$\pm$1.18  &  0.621  &
49981ce  &  811.9$\pm$24.8  &  1159.0$\pm$35.5  &  100.54$\pm$2.22  &  0.682\\
49868dk  &  1021.6$\pm$22.9  &  1100.3$\pm$24.6  &  95.83$\pm$2.39  &  0.617  &
49984ce  &  654.5$\pm$20.7  &  1107.2$\pm$35.1  &  97.62$\pm$2.09  &  0.692\\
49869ak  &  1111.5$\pm$27.9  &  1051.3$\pm$26.3  &  96.56$\pm$1.25  &  0.643  &
49985ce  &  526.3$\pm$16.1  &  1153.4$\pm$35.4  &  99.81$\pm$2.69  &  0.681\\
49869bk  &  828.2$\pm$21.0  &  1092.6$\pm$27.7  &  96.25$\pm$1.08  &  0.634  &
49986be  &  824.8$\pm$24.7  &  1096.9$\pm$32.9  &  96.95$\pm$1.11  &  0.682\\
49869ce  &  671.2$\pm$19.6  &  1042.6$\pm$30.5  &  98.57$\pm$1.98  &  0.628  &
49988ac  &  1316.6$\pm$40.1  &  1114.5$\pm$34.0  &  99.44$\pm$2.66  &  0.715\\
49869ck  &  926.6$\pm$26.8  &  1108.9$\pm$32.1  &  95.52$\pm$1.24  &  0.633  &  
50007be  &  963.7$\pm$31.1  &  1146.1$\pm$37.0  &  96.57$\pm$1.94  &  0.663\\
49869dk  &  987.1$\pm$22.2  &  1044.6$\pm$23.5  &  95.55$\pm$1.21  &  0.612  &
50051be  &  965.1$\pm$31.9  &  1271.9$\pm$42.1  &  96.21$\pm$1.03  &  0.619\\
49870ce  &  661.2$\pm$18.5  &  1041.9$\pm$29.2  &  98.05$\pm$2.95  &  0.615  &
50068be  &  1116.9$\pm$34.8  &  1245.5$\pm$38.8  &  94.59$\pm$1.91  &  0.592\\
49870dr  &  738.8$\pm$21.5  &  1130.7$\pm$32.9  &  91.73$\pm$2.26  &  0.644  &
&     &     &     &    \\
\hline
\end{tabular}\\
Notice: The columns of 'id' show ids of the spectra, 
including observational dates (JD-2400000) and corresponding 
instrument codes as listed in Dietrich et al. (1998). 
The columns of '$f_0$' give the line flux of the 
double-peaked broad H$\alpha$ directly measured from 
the observed spectra, in unit of $10^{-15}{\rm erg/s/cm^2}$, 
the uncertainty of $f_0$ is calculated by 
$f_0/f_1\times f_{1,err}$. The columns of '$f_1$' 
show the accurate flux of the H$\alpha$ collected 
from the AGNWATCH project, after necessary corrections. 
The columns of '$\sigma$' give the measured line widths 
of the double-peaked broad H$\alpha$, in unit of $\AA$. 
The columns of '$R$' show the flux ratio of the broad 
H$\alpha$ from the inner half part of the disk-like BLR 
to the total flux of the broad H$\alpha$ as discussed 
in Section 3.1.
\end{minipage}
\end{table}
%\end{landscape}

\renewcommand{\tabcolsep}{1.35mm}
\begin{table*}
\centering
\begin{minipage}{170mm}
\caption{Line Parameters of the broad H$\beta$ of mapped Seyfert 1 NGC5548}
\begin{tabular}{lll|lll|lll|lll}
\hline
id & f & $\sigma$ & id & f & $\sigma$ & id & f & $\sigma$ & id & f & $\sigma$  \\
\hline
7512m  &  71.4$\pm$4.2  &  35.66$\pm$0.26  &  7900a  &  71.4$\pm$4.2  &  
34.31$\pm$0.29  &  8513a  &  51.1$\pm$4.2  &  36.03$\pm$0.29  &  8992a  &  
67.3$\pm$4.7  &  35.88$\pm$0.25  \\ 
7517a  &  71.7$\pm$4.3  &  35.47$\pm$0.31  &  7912h  &  63.2$\pm$4.5  &  
35.59$\pm$0.30  &  8514a  &  48.3$\pm$4.6  &  35.51$\pm$0.45  &  9000a  &  
71.6$\pm$4.9  &  35.76$\pm$0.24  \\ 
7534a  &  73.2$\pm$4.3  &  34.14$\pm$0.35  &  7915a  &  63.7$\pm$4.3  &  
35.40$\pm$0.27  &  8516a  &  51.0$\pm$4.5  &  35.98$\pm$0.39  &  9008a  &  
69.2$\pm$4.3  &  35.82$\pm$0.20  \\ 
7535a  &  73.1$\pm$4.3  &  34.68$\pm$0.29  &  7929a  &  53.7$\pm$4.2  &  
36.32$\pm$0.43  &  8531h  &  57.0$\pm$4.3  &  36.74$\pm$0.41  &  9010w  &  
75.7$\pm$4.6  &  35.97$\pm$0.35  \\ 
7549m  &  78.2$\pm$4.3  &  31.86$\pm$0.25  &  7949a  &  52.2$\pm$4.1  &  
36.46$\pm$0.44  &  8636a  &  61.6$\pm$4.9  &  37.45$\pm$0.61  &  9013a  &  
70.4$\pm$4.3  &  36.20$\pm$0.21  \\ 
7560a  &  77.5$\pm$4.5  &  33.88$\pm$0.34  &  7957a  &  43.8$\pm$4.1  &  
37.76$\pm$0.43  &  8644a  &  64.1$\pm$4.8  &  37.71$\pm$0.37  &  9014w  &  
76.4$\pm$4.6  &  36.53$\pm$0.34  \\ 
7573a  &  78.2$\pm$4.2  &  34.25$\pm$0.26  &  7958a  &  43.9$\pm$4.1  &  
37.87$\pm$0.43  &  8651a  &  62.4$\pm$4.8  &  37.11$\pm$0.64  &  9017h  &  
68.7$\pm$4.6  &  36.50$\pm$0.35  \\ 
7573m  &  77.0$\pm$4.2  &  32.96$\pm$0.21  &  7971a  &  39.8$\pm$4.2  &  
38.30$\pm$0.37  &  8670a  &  58.1$\pm$4.6  &  37.04$\pm$0.48  &  9020a  &  
66.5$\pm$5.0  &  36.05$\pm$0.25  \\ 
7574m  &  76.4$\pm$4.5  &  33.59$\pm$0.27  &  7982a  &  35.4$\pm$4.1  &  
38.92$\pm$0.54  &  8676a  &  61.4$\pm$4.3  &  37.56$\pm$0.38  &  9029a  &  
66.0$\pm$4.7  &  36.03$\pm$0.29  \\ 
7575m  &  76.9$\pm$4.2  &  33.53$\pm$0.24  &  7990a  &  34.7$\pm$4.1  &  
38.22$\pm$0.26  &  8691a  &  51.2$\pm$4.2  &  37.77$\pm$0.38  &  9032h  &  
67.4$\pm$4.8  &  36.75$\pm$0.23  \\ 
7576m  &  74.3$\pm$4.2  &  32.79$\pm$0.22  &  7994a  &  37.8$\pm$4.2  &  
38.73$\pm$0.40  &  8699a  &  45.1$\pm$4.3  &  37.53$\pm$0.45  &  9034w  &  
71.3$\pm$4.7  &  36.38$\pm$0.29  \\ 
7582a  &  69.5$\pm$4.2  &  34.74$\pm$0.31  &  8007a  &  44.3$\pm$4.2  &  
38.85$\pm$0.40  &  8713a  &  33.2$\pm$4.3  &  38.51$\pm$0.58  &  9048a  & 
62.0$\pm$4.7  &  36.05$\pm$0.25  \\ 
7587m  &  72.6$\pm$4.5  &  34.32$\pm$0.42  &  8011h  &  45.1$\pm$4.3  &  
38.63$\pm$0.39  &  8720a  &  24.9$\pm$4.2  &  38.46$\pm$0.54  &  9056a  & 
65.1$\pm$4.8  &  37.01$\pm$0.25  \\ 
7589m  &  70.5$\pm$4.6  &  34.04$\pm$0.28  &  8013a  &  44.9$\pm$4.2  &  
38.06$\pm$0.27  &  8726a  &  23.7$\pm$4.2  &  39.65$\pm$0.61  &  9062a  &  
70.2$\pm$4.5  &  36.69$\pm$0.30  \\ 
7592m  &  71.4$\pm$4.6  &  33.50$\pm$0.21  &  8020a  &  49.5$\pm$4.2  &  
38.09$\pm$0.41  &  8733a  &  21.3$\pm$4.2  &  38.89$\pm$0.64  &  9064a  &  
68.9$\pm$4.6  &  36.02$\pm$0.23  \\ 
7594m  &  62.5$\pm$4.2  &  34.00$\pm$0.33  &  8028a  &  52.5$\pm$4.1  &  
37.75$\pm$0.26  &  8733h  &  23.0$\pm$4.2  &  39.04$\pm$0.56  &  9070w  &  
73.5$\pm$4.5  &  36.38$\pm$0.27  \\ 
7597m  &  63.7$\pm$4.2  &  33.58$\pm$0.24  &  8037a  &  55.5$\pm$4.2  &  
38.02$\pm$0.35  &  8742a  &  20.8$\pm$4.2  &  40.25$\pm$0.81  &  9074w  &  
71.3$\pm$4.4  &  36.42$\pm$0.92  \\ 
7601m  &  64.8$\pm$4.2  &  34.10$\pm$0.21  &  8044a  &  59.3$\pm$4.2  &  
37.66$\pm$0.28  &  8742w  &  22.3$\pm$4.1  &  37.12$\pm$1.13  &  9077w  &  
73.2$\pm$4.5  &  37.05$\pm$0.34  \\ 
7606m  &  57.9$\pm$4.2  &  31.52$\pm$0.22  &  8056h  &  60.0$\pm$4.2  &  
37.83$\pm$0.30  &  8744w  &  23.0$\pm$4.1  &  40.82$\pm$0.80  &  9078m  &  
67.3$\pm$4.3  &  36.60$\pm$0.27  \\ 
7618a  &  74.0$\pm$5.1  &  34.54$\pm$0.31  &  8061a  &  61.1$\pm$4.1  &  
37.24$\pm$0.22  &  8745w  &  23.6$\pm$4.1  &  39.09$\pm$0.76  &  9078w  &  
73.0$\pm$4.4  &  36.57$\pm$0.27  \\ 
7627a  &  77.5$\pm$5.0  &  33.64$\pm$0.41  &  8068a  &  58.7$\pm$4.2  &  
37.30$\pm$0.26  &  8746w  &  22.2$\pm$4.1  &  40.99$\pm$0.60  &  9079a  &  
65.7$\pm$4.7  &  36.24$\pm$0.29  \\ 
7642a  &  84.9$\pm$4.8  &  33.04$\pm$0.15  &  8077a  &  50.9$\pm$4.1  &  
36.88$\pm$0.24  &  8750a  &  22.0$\pm$4.1  &  39.71$\pm$0.60  &  9080m  & 
67.1$\pm$4.3  &  36.88$\pm$0.32  \\ 
7644m  &  78.1$\pm$4.2  &  32.49$\pm$0.23  &  8089h  &  45.7$\pm$4.4  &  
38.14$\pm$0.42  &  8765h  &  26.3$\pm$4.1  &  41.01$\pm$0.37  &  9083m  & 
65.5$\pm$4.2  &  36.67$\pm$0.33  \\ 
7648m  &  83.5$\pm$4.3  &  32.70$\pm$0.16  &  8089m  &  42.8$\pm$4.3  &  
37.34$\pm$0.42  &  8777w  &  28.2$\pm$4.1  &  40.22$\pm$0.59  &  9085a  &  
64.5$\pm$4.3  &  36.56$\pm$0.31  \\ 
7649a  &  90.5$\pm$4.6  &  32.95$\pm$0.21  &  8090a  &  44.7$\pm$4.1  &  
37.52$\pm$0.28  &  8778w  &  27.3$\pm$4.1  &  38.87$\pm$0.58  &  9088w  & 
66.7$\pm$4.6  &  36.83$\pm$0.36  \\ 
7652m  &  88.0$\pm$4.2  &  33.30$\pm$0.22  &  8097a  &  44.1$\pm$4.1  &  
38.39$\pm$0.36  &  8780a  &  29.4$\pm$4.1  &  40.04$\pm$0.54  &  9089w  &  
66.0$\pm$4.5  &  36.86$\pm$0.31  \\ 
7652m  &  88.8$\pm$4.2  &  33.31$\pm$0.23  &  8102h  &  46.2$\pm$4.1  & 
39.25$\pm$0.36  &  8783h  &  30.2$\pm$4.1  &  39.31$\pm$0.46  &  9090a  &  
58.8$\pm$4.4  &  36.59$\pm$0.34  \\ 
7653a  &  91.7$\pm$4.4  &  32.94$\pm$0.27  &  8128a  &  60.3$\pm$4.1  &  
38.63$\pm$0.37  &  8789a  &  37.8$\pm$4.3  &  38.26$\pm$0.41  &  9090h  &  
61.5$\pm$4.4  &  37.03$\pm$0.27  \\ 
7653a  &  94.6$\pm$4.3  &  33.08$\pm$0.26  &  8143a  &  52.9$\pm$4.2  &  
38.97$\pm$0.33  &  8796a  &  36.8$\pm$4.3  &  41.38$\pm$0.51  &  9091h  & 
60.5$\pm$4.4  &  36.93$\pm$0.29  \\ 
7654a  &  87.5$\pm$4.3  &  32.61$\pm$0.19  &  8148a  &  51.7$\pm$4.2  &  
38.82$\pm$0.45  &  8804a  &  31.2$\pm$4.1  &  38.81$\pm$0.35  &  9092h  & 
59.8$\pm$4.3  &  37.21$\pm$0.24  \\ 
7654a  &  90.0$\pm$4.2  &  32.78$\pm$0.25  &  8149a  &  52.1$\pm$4.1  &  
39.33$\pm$0.31  &  8804w  &  34.8$\pm$4.1  &  39.50$\pm$0.51  &  9099a  & 
55.7$\pm$4.4  &  36.99$\pm$0.23  \\ 
7655a  &  93.1$\pm$4.8  &  33.69$\pm$0.23  &  8151a  &  49.8$\pm$4.2  &  
39.10$\pm$0.41  &  8810h  &  28.9$\pm$4.1  &  38.45$\pm$0.46  &  9107a  &  
56.9$\pm$4.5  &  36.15$\pm$0.28  \\ 
7656m  &  88.2$\pm$4.2  &  33.91$\pm$0.22  &  8160a  &  49.4$\pm$4.2  & 
39.08$\pm$0.31  &  8817a  &  28.7$\pm$4.3  &  41.21$\pm$0.63  &  9107h  & 
59.6$\pm$4.7  &  36.44$\pm$0.29  \\ 
7657a  &  90.4$\pm$4.5  &  33.06$\pm$0.18  &  8179a  &  56.2$\pm$4.4  &  
38.52$\pm$0.42  &  8825a  &  27.3$\pm$4.1  &  39.09$\pm$0.44  &  9114a  &
63.3$\pm$4.5  &  35.41$\pm$0.30  \\ 
7657a  &  92.4$\pm$4.4  &  33.57$\pm$0.28  &  8231a  &  64.3$\pm$4.2  & 
37.08$\pm$0.26  &  8830w  &  28.7$\pm$4.1  &  38.55$\pm$0.46  &  9128a  & 
65.5$\pm$4.5  &  35.54$\pm$0.24  \\ 
7663m  &  92.1$\pm$4.6  &  33.06$\pm$0.17  &  8236a  &  64.2$\pm$4.3  & 
36.61$\pm$0.23  &  8831a  &  26.9$\pm$4.1  &  38.37$\pm$0.47  &  9135a  & 
67.8$\pm$4.5  &  36.07$\pm$0.20  \\ 
7663m  &  91.3$\pm$5.0  &  33.62$\pm$0.26  &  8252a  &  68.6$\pm$4.3  & 
35.77$\pm$0.15  &  8831w  &  30.6$\pm$4.1  &  38.75$\pm$0.36  &  9141w  & 
69.7$\pm$4.4  &  37.65$\pm$0.32  \\ 
7663m  &  95.1$\pm$4.8  &  33.35$\pm$0.22  &  8267a  &  69.3$\pm$4.5  &  
34.90$\pm$0.19  &  8833w  &  33.0$\pm$4.1  &  39.24$\pm$0.51  &  9142a  & 
62.2$\pm$4.7  &  36.08$\pm$0.82  \\ 
7664m  &  90.1$\pm$4.4  &  33.45$\pm$0.29  &  8275a  &  68.6$\pm$4.3  &  
35.03$\pm$0.21  &  8834w  &  32.1$\pm$4.1  &  38.40$\pm$0.44  &  9149a  & 
66.1$\pm$4.5  &  35.76$\pm$0.21  \\ 
7665m  &  89.7$\pm$4.2  &  33.18$\pm$0.21  &  8280a  &  68.3$\pm$4.3  &  
35.06$\pm$0.26  &  8835w  &  34.0$\pm$4.1  &  38.73$\pm$0.44  &  9156a  &
63.9$\pm$4.4  &  36.97$\pm$0.25  \\ 
7666m  &  88.5$\pm$4.2  &  33.38$\pm$0.18  &  8287a  &  69.5$\pm$4.3  & 
35.19$\pm$0.18  &  8836w  &  33.4$\pm$4.1  &  38.76$\pm$0.46  &  9156w  & 
63.8$\pm$4.4  &  36.68$\pm$0.34  \\ 
7668m  &  88.1$\pm$4.2  &  33.94$\pm$0.24  &  8294a  &  67.7$\pm$5.0  &  
34.79$\pm$0.32  &  8837h  &  29.9$\pm$4.1  &  38.58$\pm$0.36  &  9157w  & 
63.3$\pm$4.3  &  37.02$\pm$0.34  \\ 
7678a  &  86.7$\pm$4.3  &  33.86$\pm$0.29  &  8310a  &  65.3$\pm$4.3  &  
35.15$\pm$0.27  &  8837w  &  32.0$\pm$4.1  &  38.62$\pm$0.48  &  9158w  & 
60.0$\pm$4.2  &  36.32$\pm$0.33  \\ 
7711a  &  78.6$\pm$4.5  &  35.04$\pm$0.39  &  8323a  &  71.3$\pm$4.4  &  
35.82$\pm$0.28  &  8839a  &  30.3$\pm$4.2  &  39.19$\pm$0.38  &  9159w  &  
60.9$\pm$4.2  &  37.08$\pm$0.25  \\ 
7713m  &  68.6$\pm$4.1  &  34.41$\pm$0.25  &  8338a  &  72.4$\pm$4.3  &  
34.78$\pm$0.21  &  8848a  &  34.7$\pm$4.3  &  37.35$\pm$0.41  &  9163a  & 
57.3$\pm$4.3  &  36.52$\pm$0.22  \\ 
7725a  &  84.6$\pm$4.6  &  35.43$\pm$0.38  &  8344a  &  72.6$\pm$4.6  &  
36.24$\pm$0.24  &  8858h  &  39.5$\pm$4.1  &  36.63$\pm$0.29  &  9166h  & 
59.0$\pm$4.4  &  36.99$\pm$0.24  \\ 
7746m  &  71.1$\pm$4.4  &  36.53$\pm$0.38  &  8351a  &  72.0$\pm$4.2  &  
35.81$\pm$0.31  &  8858w  &  46.0$\pm$4.2  &  38.02$\pm$0.29  &  9169a  & 
57.8$\pm$4.2  &  36.35$\pm$0.23  \\ 
7749m  &  60.5$\pm$4.1  &  35.74$\pm$0.26  &  8352a  &  69.4$\pm$4.3  &  
35.25$\pm$0.30  &  8861w  &  49.0$\pm$4.2  &  38.60$\pm$0.39  &  9176a  &  
60.2$\pm$4.3  &  36.38$\pm$0.25  \\ 
7754m  &  66.0$\pm$4.1  &  35.68$\pm$0.38  &  8365a  &  58.1$\pm$4.2  &  
36.05$\pm$0.33  &  8862a  &  42.6$\pm$4.3  &  37.25$\pm$0.36  &  9182h  & 
63.3$\pm$4.2  &  36.29$\pm$0.35  \\ 
7757m  &  64.1$\pm$4.5  &  35.58$\pm$0.34  &  8378a  &  57.2$\pm$4.3  &  
37.18$\pm$0.41  &  8862w  &  45.7$\pm$4.2  &  36.25$\pm$0.24  &  9183a  & 
63.1$\pm$4.4  &  36.91$\pm$0.23  \\ 
7758m  &  60.2$\pm$4.3  &  36.72$\pm$0.40  &  8386a  &  57.4$\pm$4.5  &  
37.07$\pm$0.35  &  8867h  &  42.9$\pm$4.3  &  38.21$\pm$0.35  &  9190a  &  
64.6$\pm$4.5  &  36.64$\pm$0.22  \\ 
7759m  &  60.7$\pm$4.1  &  35.71$\pm$0.34  &  8393a  &  58.4$\pm$4.2  &  
36.99$\pm$0.35  &  8869a  &  41.0$\pm$4.5  &  37.55$\pm$0.42  &  9196h  & 
66.9$\pm$4.4  &  36.49$\pm$0.37  \\ 
7765m  &  58.2$\pm$4.1  &  35.49$\pm$0.33  &  8400a  &  59.1$\pm$4.3  &  
36.94$\pm$0.27  &  8876a  &  45.4$\pm$4.3  &  37.92$\pm$0.35  &  9197a  & 
64.8$\pm$4.5  &  36.25$\pm$0.24  \\ 
7767m  &  59.1$\pm$4.1  &  36.10$\pm$0.27  &  8414a  &  48.7$\pm$4.2  &  
36.77$\pm$0.38  &  8883a  &  44.8$\pm$4.5  &  37.53$\pm$0.32  &  9205a  & 
66.2$\pm$4.3  &  37.18$\pm$0.32  \\ 
7777h  &  66.3$\pm$4.6  &  36.02$\pm$0.30  &  8421a  &  43.8$\pm$4.2  &  
34.96$\pm$0.28  &  8886h  &  43.8$\pm$4.3  &  37.64$\pm$0.32  &  9211a  &  
63.1$\pm$4.5  &  36.78$\pm$0.23  \\ 
7797h  &  78.3$\pm$4.9  &  35.24$\pm$0.36  &  8431a  &  49.4$\pm$4.4  &  
36.19$\pm$0.38  &  8889a  &  45.6$\pm$4.5  &  37.08$\pm$0.35  &  9212h  & 
65.6$\pm$4.8  &  37.59$\pm$0.26  \\ 
7809h  &  80.0$\pm$4.7  &  34.46$\pm$0.25  &  8455a  &  58.8$\pm$4.2  &  
36.11$\pm$0.31  &  8898a  &  49.2$\pm$4.3  &  38.17$\pm$0.25  &  9240a  &  
58.7$\pm$4.5  &  38.15$\pm$0.23  \\ 
7868a  &  77.5$\pm$4.4  &  35.77$\pm$0.27  &  8456a  &  58.2$\pm$4.2  &  
36.55$\pm$0.40  &  8898h  &  46.2$\pm$4.4  &  37.18$\pm$0.37  &  9240h  &  
60.5$\pm$4.7  &  38.48$\pm$0.25  \\ 
7884a  &  79.1$\pm$4.2  &  34.55$\pm$0.25  &  8460a  &  53.5$\pm$4.2  &  
35.69$\pm$0.41  &  8954a  &  57.6$\pm$4.8  &  37.06$\pm$0.32  &  9242h  & 
60.2$\pm$4.6  &  38.13$\pm$0.30  \\ 
7891a  &  80.0$\pm$4.3  &  35.00$\pm$0.37  &  8474m  &  43.5$\pm$4.3  &  
37.56$\pm$0.41  &  8967a  &  59.2$\pm$4.6  &  36.85$\pm$0.28  &  9243h  &  
59.4$\pm$4.7  &  38.23$\pm$0.27  \\ 
7895h  &  75.6$\pm$4.4  &  35.46$\pm$0.38  &  8477m  &  41.1$\pm$4.1  &  
37.27$\pm$0.42  &  8973a  &  64.0$\pm$5.1  &  38.77$\pm$0.73  &  9255h  &  
58.7$\pm$4.7  &  38.79$\pm$0.27  \\ 
7899a  &  72.4$\pm$4.7  &  34.88$\pm$0.41  &  8512a  &  51.8$\pm$4.3  &  
36.27$\pm$0.29  &  8981a  &  64.9$\pm$4.6  &  36.27$\pm$0.17  &    &  &   \\ 
\hline
\end{tabular}\\
Notice: The columns of 'id' show the observational date 
(JD-2450000) and instrument codes. The columns of '$f$' 
list the measured line flux of the broad H$\beta$ in unit 
of $10^{-14}{\rm erg/s/cm^2}$. The columns of '$\sigma$' 
show the second moment of the broad H$\beta$ in unit of $\AA$.
\end{minipage}
\end{table*}

\clearpage

\label{lastpage}

\begin{thebibliography}{   }
\bibitem[\protect\citeauthoryear{Bachev}{1999}]{ba9}
Bachev, R., 1999, A\&A, 348, 71
\bibitem[\protect\citeauthoryear{Barth et al.}{2011}]{bn11}
Barth A. J., Nguyen M. L., Malkan M. A., Filippenko A. V., Li W. D.,  
     et al., 2011, ApJ, 732, 121
\bibitem[\protect\citeauthoryear{Begelman et al.}{1980}]{beg80}
Begelman M. C., Blandford R. D. \& Rees M. J., 1980, Nature, 287, 307
\bibitem[\protect\citeauthoryear{Bennert et al.}{2011}]{ba11}
Bennert N., Auger M. W., Treu T., Woo J.-H., Malkan M. A., 2011, 
    ApJ, 726, 59
\bibitem[\protect\citeauthoryear{Bentz et al.}{2006}]{ben06}
Bentz M. C., Peterson B. M., Pogge R. W., Vestergaard M., 
     Onken C. A., 2006, ApJ, 644, 133
\bibitem[\protect\citeauthoryear{Bentz et al.}{2007}]{ben07}
Bentz M. C., Denney K. D., Cackett E. M., Dietrich M., Fogel J. K. J., 
    et al., 2007, ApJ, 662, 205
\bibitem[\protect\citeauthoryear{Bentz et al.}{2009}]{bw09}
Bentz M. C., Walsh J. L., Barth A. J., Baliber N., Bennert V. N., 
    et al., 2009, ApJ, 705, 199
\bibitem[\protect\citeauthoryear{Bentz et al.}{2010}]{bw10}
Bentz M. C., Walsh J. L., Barth A. J., Yoshii Y., Woo J. H., et al., 
    2010, ApJ, 716, 993
\bibitem[\protect\citeauthoryear{Blandford \& Mckee}{1982}]{bm82}
Blandford R. D. \& Mckee C. F., 1982, ApJ, 255, 419
\bibitem[\protect\citeauthoryear{Boroson \& Laue}{2009}]{bl09}
Boroson T. A. \& Lauer T. R., 2009, Nature, 458, 53
\bibitem[\protect\citeauthoryear{Burbidge \& Burbidge}{1971}]{bb71}
Burbidge E. M. \& Burbidge G. R., 1971, ApJ, 163, L21
\bibitem[\protect\citeauthoryear{Cao}{2005}]{cao05}
Cao X. W., 2005, ApJ, 631, L101
\bibitem[\protect\citeauthoryear{Cao \& Wang}{2006}]{cw06}
Cao X. W. \& Wang T. G., 2006, ApJ, 652, 112
\bibitem[\protect\citeauthoryear{Chen \& Halpern}{1989}]{ch89a}
Chen K. Y. \& Halpern J. P., 1989, ApJ, 344, 115
\bibitem[\protect\citeauthoryear{Chen et al.}{1989}]{ch89b}
Chen K. Y., Halpern J. P. \& Filippenko A. V., 1989, ApJ, 339, 742
\bibitem[\protect\citeauthoryear{Chornock et al.}{2010}]{ch10}
Chornock R., Bloom J. S., Cenko S. B., Filippenko A. V., Silverman J. M.,
    Hicks M. D., Lawrence K. J., Mendez A. J., Rafelski M., Wolfe, A. M.,
    2010, ApJ, 709, 39
\bibitem[\protect\citeauthoryear{Collin-Souffrin}{1987}]{cs87}
Collin-Souffrin S. 1987, A\&A, 179, 60
\bibitem[\protect\citeauthoryear{Collin et al.}{2006}]{co06}
Collin S., Kawaguchi T., Peterson B. M., Vestergaard M., 2006, A\&A, 456, 75
\bibitem[\protect\citeauthoryear{Davis \& Laor}{2011}]{dl11}
Davis S. W. \& Laor A., 2011, ApJ, 728, 98
\bibitem[\protect\citeauthoryear{Denney et al.}{2006}]{db06}
Denney K. D., Bentz M. C., Peterson B. M., Pogge R. W., Cackett E. M., et al., 
    2006, ApJ, 653, 152
\bibitem[\protect\citeauthoryear{Denney et al.}{2009}]{dp09}
Denney K. D., Peterson B. M., Pogge R. W., Adair A., Atlee D. W., et al., 
    2009, ApJL, 704, 80
\bibitem[\protect\citeauthoryear{Denney et al.}{2010}]{dp10}
Denney K. D., Peterson B. M., Pogge R. W., Adair A., Atlee D. W., 
    et al., 2010, ApJ, 721, 715
\bibitem[\protect\citeauthoryear{Dietrich et al.}{1993}]{die93}
Dietrich, M., Kollatschny, W., Peterson, B. M., Bechtold, J., 
     Bertram R., et al., 1993, ApJ, 408, 416
\bibitem[\protect\citeauthoryear{Dietrich et al.}{1998}]{di98}
Dietrich M., Peterson B. M., Albrecht P., Altmann M., Barth A. J., et al.,
    1998, ApJS, 115, 185
\bibitem[\protect\citeauthoryear{Dietrich et al.}{2012}]{di12}
Dietrich M., Peterson B. M., Grier C. J., Bentz M .C., Eastman J., et al., 
    2012, ApJ, 757, 53D
\bibitem[\protect\citeauthoryear{Down et al.}{2010}]{dr10}
Down E. J., Rawlings S., Sivia D. S., Baker J. C., 2010, MNRAS, 401, 633
\bibitem[\protect\citeauthoryear{Dumont \& Collin-Souffrin}{1990}]{dc90}
Dumont A. M. \& Collin-Souffrin S. 1990, A\&A, 229, 313
\bibitem[\protect\citeauthoryear{Eracleous \& Halpern}{1994}]{eh94}
Eracleous M. \& Halpern J. P., 1994, ApJS, 90, 1 
\bibitem[\protect\citeauthoryear{Eracleous et al.}{1995}]{era95}
Eracleous M., Livio M., Halpern J. P., Storchi-Bergmann T., 1995, ApJ, 438, 610
\bibitem[\protect\citeauthoryear{Eracleous et al.}{1997}]{era97}
Eracleous M., Halpern J. P., Gilbert A. M., Newman J. A., Filippenko A. V.,
    1997, ApJ, 490, 216
\bibitem[\protect\citeauthoryear{Eracleous \& Halpern}{2003}]{eh03}
Eracleous M. \& Halpern J. P., 2003, ApJ, 599, 886
\bibitem[\protect\citeauthoryear{Flohic \& Eracleous}{2008}]{fe08}
Flohic H. M. L. G. \& Eracleous M., 2008, ApJ, 686, 138
\bibitem[\protect\citeauthoryear{Fromerth \& Melia}{2000}]{fm00}
Fromerth M. J. \& Melia F., 2000, ApJ, 533, 172
\bibitem[\protect\citeauthoryear{Gaskell}{1988}]{gas88}
Gaskell C. M., 1988, ApJ, 325, 114
\bibitem[\protect\citeauthoryear{Gaskell}{1996}]{gas96}
Gaskell M., 1996, ApJ, 464, 107
\bibitem[\protect\citeauthoryear{Gaskell}{2010}]{gas10}
Gaskell C. M., 2010, Nature, 463, 1
\bibitem[\protect\citeauthoryear{Gezari et al.}{2007}]{gh07}
Gezari S., Halpern J. P., Eracleous M., 2007, ApJ, 169, 167
\bibitem[\protect\citeauthoryear{Greene \& Ho}{2004}]{gh04}
Greene J. E. \& Ho L. C., 2004, ApJ, 610, 722
\bibitem[\protect\citeauthoryear{Greene \& Ho}{2005}]{gh05}
Greene J. E. \& Ho L. C., 2005, ApJ, 630, 122
\bibitem[\protect\citeauthoryear{Greene et al.}{2010}]{gh10}
Greene J. E., Hood C. E., Barth A. J., Bennert V. N., Bentz M. C., 
    et al., 2010, ApJ, 723, 409
\bibitem[\protect\citeauthoryear{Hartnoll \& Blackman}{2000}]{hb00}
Hartnoll S. A. \& Blackman E. G., 2000, MNRAS, 317, 880
\bibitem[\protect\citeauthoryear{Hartnoll \& Blackman}{2002}]{hb02}
Hartnoll S. A. \& Blackman E. G., 2002, MNRAS, 332, L1
\bibitem[\protect\citeauthoryear{Karas et al.}{2001}]{kar01}
Karas V., Martocchia A. \& Subr L., 2001, PASJ, 53, 189
\bibitem[\protect\citeauthoryear{Kaspi et al.}{1996}]{kas96}
Kaspi S., Smith P. S.,  Maoz D.,  Netzer H., Jannuzi B. T., 
    1996, ApJL, 471, 75
\bibitem[\protect\citeauthoryear{Kaspi et al.}{2000}]{kas00}
Kaspi S., Smith P. S., Netzer H., Maoz D., Jannuzi B. T., Giveon U., 2000,
ApJ, 533, 631
\bibitem[\protect\citeauthoryear{Kaspi et al.}{2005}]{kas05}
Kaspi S., Maoz D., Netzer H., Peterson B. M., Vestergaard M.,
Jannuzi B. T., 2005, ApJ, 629, 61
\bibitem[\protect\citeauthoryear{Kaspi et al.}{2007}]{kas07}
Kaspi S., Brandt W. N., Maoz D., Netzer H., Schneider D. P., Shemmer O., 2007, 
       ApJ, 659, 997
\bibitem[\protect\citeauthoryear{Kelly \& Bechtold}{2007}]{kb07}
Kelly B. C. \& Bechtold J., 2007, ApJS, 168, 1
\bibitem[\protect\citeauthoryear{Kollatschny \& Zetzl1}{2011}]{kz11}
Kollatschny W. \& Zetzl1 M., 2011, Nature, 470, 366
\bibitem[\protect\citeauthoryear{Korista et al.}{1995}]{ka95}
Korista K. T., Alloin D., Barr P., Clavel J., Cohen R. D., 
     et al., 1995, ApJS, 97, 285
\bibitem[\protect\citeauthoryear{Krause et al.}{2011}]{kb11}
Krause M., Burkert A., Schartmann M., 2011, MNRAS, 411, 550
\bibitem[\protect\citeauthoryear{Lauer \& Boroson}{2009}]{lb09}
Lauer T. R. \& Boroson T. A. 2009, ApJ, 703, 930
\bibitem[\protect\citeauthoryear{Lewis et al.}{2010}]{lew10}
Lewis K. T., Eracleous M., Storchi-Bergmann T., 2010, ApJS, 187, 416
\bibitem[\protect\citeauthoryear{Livio \& Xu}{1997}]{lx97}
Livio M. \& Xu C., 1997, ApJ, 478, L63
\bibitem[\protect\citeauthoryear{Luo et al.}{2009}]{lbs09}
Luo B., Brandt W. N., Silverman J. D., Strateva I. V., Bauer F. E., et al.,
    2009, ApJ, 695, 1227
\bibitem[\protect\citeauthoryear{Marziani et al.}{2003}]{mar03}
Marziani P., Sulentic J. W., Zamanov R., Calvani M., Dultzin-Hacyan D.,
   Bachev R., Zwitter T., 2003, ApJS, 145, 199
\bibitem[\protect\citeauthoryear{Manmoto}{2000}]{ma00}
Manmoto T., 2000, ApJ, 534, 734
\bibitem[\protect\citeauthoryear{Netzer \& Marziani}{2010}]{nm10}
Netzer H. \& Marziani P., 2010, ApJ, 724, 318
\bibitem[\protect\citeauthoryear{Oke}{1987}]{oke87}
Oke J. B., IN: Superluminal radio sources; Proceedings of the Workshop, 
   Pasadena, CA, Oct. 28-30, 1986 (A88-39751 16-90). Cambridge and New York, 
   Cambridge University Press, 1987, p. 267-272.
\bibitem[\protect\citeauthoryear{Onken et al.}{2004}]{on04}
Onken C. A., Ferrarese L., Merritt D., Peterson B. M., Pogge R. W.,
   Vestergaard M., Wandel A., 2004, ApJ, 615, 645
\bibitem[\protect\citeauthoryear{Park et al.}{2004}]{pa12}
Park D., Woo J. H., Treu T., Barth A. J., Bentz M. C., et al., 2012, ApJ, 
    747, 30
\bibitem[\protect\citeauthoryear{Perez et al.}{1988}]{pe88}
Perez E., Penston M. V., Tadhunter C., Mediavilla E., Moles M., 1988, 
    MNRAS, 230, 353
\bibitem[\protect\citeauthoryear{Peterson et al.}{1998}]{pe98}
Peterson B. M., Wanders I., Horne K., Collier S., 1998, PASP, 110, 660
\bibitem[\protect\citeauthoryear{Peterson \& Wandel}{1999}]{pw99}
Peterson B. M. \& Wandel A., 1999, ApJ, 521, L95
\bibitem[\protect\citeauthoryear{Peterson et al.}{1991}]{pe91}
Peterson B. M., Balonek T. J., Barker E. S., Bechtold J., Bertram R.,et al.,
    1991, ApJ, 368, 119
\bibitem[\protect\citeauthoryear{Peterson et al.}{1992}]{pe92}
Peterson B. M., Alloin D., Axon D., Balonek T. J., Bertram R., et al.,
    1992, ApJ, 392, 470
\bibitem[\protect\citeauthoryear{Peterson}{1993}]{pe93}
Peterson, B. M., 1993, PASP, 105, 247
\bibitem[\protect\citeauthoryear{Peterson et al.}{1994}]{pb94}
Peterson B. M., Berlind P., Bertram R., Bochkarev N. G., Bond D., et al.,
    1994, ApJ, 425, 622
\bibitem[\protect\citeauthoryear{Peterson et al.}{1999}]{pe99}
Peterson B. M., Barth A. J., Berlind P., Bertram R., Bischoff K., 
    et al., 1999, ApJ, 510, 659
\bibitem[\protect\citeauthoryear{Peterson et al.}{2002}]{pe02}
Peterson B. M., Berlind P., Bertram R., Bischoff K., Bochkarev N. G., 
    et al., 2002, ApJ, 581, 197
\bibitem[\protect\citeauthoryear{Peterson et al.}{2004}]{pe04}
Peterson B. M., Ferrarese L., Gilbert K. M., Kaspi S., et al., 2004,
      ApJ, 613, 682
\bibitem[\protect\citeauthoryear{Peterson}{2010}]{pe10}
Peterson B. M., 2010, Co-Evolution of Central Black Holes and Galaxies,
     Proceedings of the International Astronomical Union, IAU Symposium,
     Volume 267, p. 151-160
\bibitem[\protect\citeauthoryear{Popovic et al.}{2008}]{ps08}
Popovic L. C., Shapovalova A. I., Chavushyan V. H., Ilic D., Burenkov A. N.,
       Mercado A., Bochkarev N. G., 2008, PASJ, 60, 1
\bibitem[\protect\citeauthoryear{Popovic et al.}{2011}]{ps11}
Popovic L. C., Shapovalova A. I., Ilic D., Kovacevic A., Kollatschny W., 
       Burenkov A. N., Chavushyan V. H., Bochkarev N. G., Leon-Tavares J., 
       2011, A\&A, 528, 130
\bibitem[\protect\citeauthoryear{Press et al.}{1992}]{pr92}
Press W. H., Teukolsky S. A., Vetterling W. T., Flannery B. P., 1992,
    'Numerical Recipes in Fortran 77', Second Edition, published by the
     Press Syndicate of the University of Cambridge, ISBN 0-521-43064-X, P340
\bibitem[\protect\citeauthoryear{Rafiee \& Hall}{2011}]{rh11}
Rafiee A., \& Hall P. B., 2011, ApJS, 194, 42
\bibitem[\protect\citeauthoryear{Schimoia et al.}{2012}]{ss12}
Schimoia J. S., Storchi-Bergmann T., Nemmwn R. S., Winge C., Eracleous M.,
    2012, ApJ, 748, 145
\bibitem[\protect\citeauthoryear{Sergeev et al.}{2007}]{sd07}
Sergeev S. G., Doroshenko V. T., Dzyuba S. A., Peterson B. M., Pogge R. W.,
          Pronik V. I., 2007, ApJ, 668, 708
\bibitem[\protect\citeauthoryear{Sergeev et al.}{2011}]{sk11}
Sergeev S. G., Kilmanov S. A., Doroshenko V. T., Efimov Y. S., Nazarov S. V., 
          Pronik V. I., 2011, MNRAS, 410, 1877          
\bibitem[\protect\citeauthoryear{Shapovalova et al.}{2001}]{sh01}
Shapovalova A. I., Burenkov A. N., Carrasco L., Chavushyan V. H.,
     Doroshenko V. T., et al., 2001, A\&A, 376, 775
\bibitem[\protect\citeauthoryear{Shapovalova et al.}{2004}]{sd04}
Shapovalova A. I., Doroshenko V. T., Bochkarev N. G., Burenkov A. N.,
    Carrasco L., et al., 2004, 422, 925
\bibitem[\protect\citeauthoryear{Shapovalova et al.}{2010}]{sh10}
Shapovalova A. I., Popovic L. C., Burenkov A. N., Chavushyan V. H., Ilic D.,
     et al., 2010, A\&A, 517, 42
\bibitem[\protect\citeauthoryear{Shen \& Liu}{2012}]{sl12}
Shen Y. \& Liu X., 2012, ApJ, 753, 125
\bibitem[\protect\citeauthoryear{Sluse et al.}{2011}]{ss11}
Sluse D., Schmidt R., Courbin F., Hutsemekers D., Meylan G.,
   Eigenbrod A., Anguita T., Agol E., Wambsganss J., 2011, A\&A, 528, 100
\bibitem[\protect\citeauthoryear{Sulentic et al.}{2000}]{su00}
Sulentic J. W., Marziani P., Dultzin-Hacyan D., 2000, ARA\&A, 38, 521
\bibitem[\protect\citeauthoryear{Storchi-Bergmann et al.}{2003}]{sn03}
Storchi-Bergmann T., Nemmen da S. R., Eracleous M., Halpern J. P.,
     Wilson A. S., Filippenko A. V., Ruiz M. T., Smith R. C., Nagar N. M.,
     1997, ApJ, 489, 8
\bibitem[\protect\citeauthoryear{Strateva et al.}{2003}]{sb03}
Strateva I. V., Strauss M. A., Hao L., Schlegel D. J., Hall P. B., et al., 
   2003, AJ, 126, 1720
\bibitem[\protect\citeauthoryear{Strateva et al.}{2006}]{sb06}
Strateva I. V., Brandt W. N., Eracleous M., Schneider D. P., Chartas, G. 
     2006, ApJ, 651, 749
\bibitem[\protect\citeauthoryear{Strateva et al.}{2008}]{sb08}
Strateva I. V., Brandt W. N., Eracleous M., Garmire G., 2008, ApJ, 687, 869
\bibitem[\protect\citeauthoryear{Tran}{2010}]{tr10}
Tran H. D., 2010, ApJ, 711, 1174
\bibitem[\protect\citeauthoryear{van Groningen \& Wanders}{1992}]{vw92}
van Groningen, E. \& Wanders, I., 1992, PASP, 104, 700
\bibitem[\protect\citeauthoryear{Veilleux \& Zheng}{1991}]{vz91}
Veilleux S. \& Zheng W., 1991, ApJ, 377, 89
\bibitem[\protect\citeauthoryear{Vestergaard}{2002}]{v02}
Vestergaard M., 2002, ApJ, 571, 733
\bibitem[\protect\citeauthoryear{Wandel et al.}{1999}]{wan99}
Wandel A., Peterson B. M., \& Malkan M. A., 1999, ApJ, 526, 579
\bibitem[\protect\citeauthoryear{Wanders \& Peterson}{1996}]{wp96}
Wanders I. \& Peterson B. M., 1996, ApJ, 466, 174
\bibitem[\protect\citeauthoryear{Wang \& Zhang}{2003}]{wz03}
Wang T. G., \& Zhang X. G., 2003, MNRAS, 340, 793
\bibitem[\protect\citeauthoryear{Williams}{1980}]{wi80}
Williams R. E., 1980, ApJ, 235, 939
\bibitem[\protect\citeauthoryear{Wu et al.}{2004}]{wu08}
Wu S. M., Wang T. G., Dong X. B., 2008, MNRAS, 389, 213
\bibitem[\protect\citeauthoryear{Wu et al.}{2004}]{ww04}
Wu X. B., Wang R., Kong M. Z., Liu F.K., Han J. L., 2004, A\&A, 424, 793
\bibitem[\protect\citeauthoryear{Zhang et al.}{2007}]{zh07}
Zhang X. G., Dultzin D., Wang T. G., 2007, MNRAS, 377, 1215
\bibitem[\protect\citeauthoryear{Zhang}{2011a}]{zh11a}
Zhang X. G., 2011a, MNRAS, 416, 2857
\bibitem[\protect\citeauthoryear{Zhang}{2011b}]{zh11b}
Zhang X. G., 2011b, ApJ, 741, 104
\bibitem[\protect\citeauthoryear{Zheng et al.}{1990}]{zh90}
Zheng W., Sulentic J. W., Binette L., 1990, ApJ, 365, 115
\bibitem[\protect\citeauthoryear{Zheng et al.}{1991}]{zh91}
Zheng W., Veilleux S., Grandi S. A., 1991, ApJ, 381, 418
\end{thebibliography}
\end{document}